\documentclass[aps,prb,twocolumn,superscriptaddress,floatfix]{revtex4-2}

\usepackage{amsmath, amssymb, graphicx}
\usepackage[colorlinks=true,citecolor=blue,urlcolor=blue,linkcolor=blue]{hyperref}
\usepackage{xcolor}
\usepackage{braket}
\usepackage[utf8]{inputenc}
\usepackage{bm}

\usepackage{booktabs}

\begin{document}

\title{Quantum Mott semimetal in a one-dimensional Hubbard model}

\author{Boran Zhou}
\affiliation{Department of Physics and Astronomy, Johns Hopkins University, Baltimore, Maryland 21218, USA}
\author{Taige Wang}
\affiliation{Department of Physics, Harvard University, Cambridge, MA 02138, USA}
\affiliation{Materials Research Laboratory, Massachusetts Institute of Technology, Cambridge, MA 02139, USA}
\author{Ya-Hui Zhang}
\affiliation{Department of Physics and Astronomy, Johns Hopkins University, Baltimore, Maryland 21218, USA}

\date{\today}
\begin{abstract}
Mott physics in topological bands has recently attracted considerable attention, particularly in the context of twisted bilayer graphene (TBG). However, the essential ingredients for stabilizing this physics remain unclear. Here, we demonstrate a quantum Mott semimetal phase as the ground state within a one-dimensional spinful Hubbard model featuring only one orbital per unit cell, protected by inversion and particle-hole symmetries. We start from a two-orbital model where a localized $f_A$ orbital on the A sublattice hybridizes with a delocalized $f_B$ orbital on the B sublattice. Projecting the $f_A$ orbital Hubbard $U$ onto the active flat band yields a lattice model with Wannier orbitals centered on the B sublattice. Similar to TBG, a momentum-space scale $k_*$ emerges, setting the interaction range in the projected model to $1/k_*$. While the ground state is ferromagnetic with only the Hubbard $U$, introducing an inter-site antiferromagnetic spin coupling $J$ stabilizes a Mott semimetal$^*$ phase with a central charge $c=3$. Using exact diagonalization (ED) and density matrix renormalization group (DMRG) methods, we show that this phase hosts a spinful Dirac fermion coexisting with a neutral spin mode---analogous to the fractionalized Fermi liquid (FL$^*$) phase in higher dimensions. Furthermore, breaking particle-hole (PH) symmetry via dispersion transforms the Mott semimetal into a Mott insulator, which is separated from a distinct Mott insulating phase by a continuous transition with a polarization jump of $1/2$. Our work provides the first unbiased evidence of a Mott semimetal ground state and demonstrates that this 1D model captures some essential aspects of TBG physics, despite lacking a Wannier obstruction.
\end{abstract}

\maketitle

\emph{Introduction.}---The interplay between strong correlations and band topology
is a central theme in the physics of moir\'e
materials~\cite{andrei2021marvels,mak2022semiconductor,Balents2020,Cao2018}.
In twisted bilayer graphene (TBG),
experiments reveal signatures of Mott insulating states at integer
fillings~\cite{rozen2021entropic,saito2021isospin,Wong2020,Zondiner2020}, 
coexisting with topologically nontrivial band
structures and quantum Hall physics~\cite{sharpe2019emergent,serlin2020intrinsic,PhysRevX.10.031034,PhysRevX.11.041063,PhysRevLett.127.027601,PhysRevLett.128.156401,PhysRevLett.124.166601,PhysRevResearch.1.033126,Xie2021}. 
While quantum Hall ferromagnets within the Hartree Fock framework
provide a natural starting point for correlated insulators in Chern
bands~\cite{zhang2019nearly,repellin2020ferromagnetism},
the experimental evidence for local moments and
``Mottness''~\cite{rozen2021entropic,saito2021isospin}
suggests physics beyond the Slater determinant picture in twisted bilayer graphene (TBG).
This has motivated the proposal of Mott semimetal~\cite{PhysRevX.15.021087,hu2025projectedsolvabletopologicalheavy,zhao2025ancilla,ledwith2025exotic,kts3-81nk,vituri2026controlledloopexpansiontopological,wei2026lifetimespectralfunctiontopological,nosov2026controlledexpansioncorrelatedelectrons} where a gapless Dirac fermion coexists with local moments, as well as the symmetric topological Mott insulators
(STMIs)~\cite{zhou2025STMI}, which realize quantized Hall responses through
interaction-driven mechanisms that cannot be smoothly connected to any
Slater determinant.

 Many studies of the Mott semimetal phase focused on the finite temperature regime~\cite{vituri2026controlledloopexpansiontopological,nosov2026controlledexpansioncorrelatedelectrons,PhysRevX.15.021087,hu2025projectedsolvabletopologicalheavy} where the local moments are thermally fluctuating. In contrast to this thermal Mott semimetal, a quantum Mott semimetal at zero temperature\cite{zhao2025ancilla,kts3-81nk} is more non-trivial because the degeneracy of the local moments needs to be lifted to form a coherent quantum state. Given the ``identical particle'' principle in quantum mechanics, it is not clear at all whether a Mott semimetal phase with both gapless itinerant electrons and local moments can emerge from electrons within a one-band model. In this work we will provide the first unbiased establishment of such a phase in one dimension.

We consider a one dimensional spinful model with a flat topological band protected by the inversion symmetry, similar to the Su-Schrieffer-Heeger (SSH) chain~\cite{su1980soliton}.  While the flat band has a well-defined Wannier orbital on a B sublattice, its density is mainly from a f orbital on A sublattice at the center of a nearest neighbor bond of B sites.  We project the Hubbard $U$ of this f orbital to the flat band and obtain a spinful Hubbard model with only the B sublattice, but the Hubbard $U$ is non-local and has a range of $1/k_*$, where $k_*$ is the size of the non-trivial momentum space region of non-trivial quantum geometry. 

With only the Hubbard $U$, the ground state is spin polarized. Remarkably, an antiferromagnetic exchange interaction on the $A$ sublattice can melt this ferromagnet into a spin-singlet ground state. The resulting phase turns out to be a C1S2 phase with one charge mode and two spin modes. It can be understood in a two component picture: in the charge sector we have a spinful Dirac fermion at $k=0$, and we have another gapless neutral spin mode from the local moments.  The physics is similar to the FL* phase discussed in higher dimension~\cite{PhysRevLett.90.216403}. We need to emphasize that here the itinerant electrons and the local moments are originating from the same electrons in the one-orbital model, which is different from the C1S2 phase in a Kondo lattice model\cite{rende2026transformer}. Also, a composite operator $\psi_{\sigma;k}$ needs to emerge and hybridizes with the single electron operator $c_{\sigma;k}$ together to form the Dirac crossing at $k=0$ for each spin. We introduce a low energy effective theory and interpret this emergent $\psi$ fermion as a composite fermion with electron bound to a particle-hole excitation.  At $k=0$, the parity quantum number of $\psi_k$ is different from that of $c_k$ and then the hybridization is enforced to be $k \sigma_x$ by the inversion symmetry in the resulting emergent two-band model. Meanwhile we need the PH symmetry to forbid the Dirac mass term $m\tau^z$.

We further show that breaking PH symmetry by adding an additional hopping term drives the system into a Mott insulating phase labeled as C0S1. However, it is still distinct to the standard  Mott insulator by a non-trivial many-polarization  $\mathbb P=\frac{1}{2}$~\cite{PhysRevLett.80.1800,PhysRevB.108.235150}. We find that increasing the on-site Hubbard $U$ drives a transition between two C0S1 Mott insulators with a jump of $\mathbb P$ by $\frac{1}{2}$.  The critical point turns out to be the usual C1S1 Luttinger liquid phase. We also note that in the Mott insulator with $\mathbb P=\frac{1}{2}$, there is Friedel oscillation with momentum $q=\pi$, indicating the non-localized nature of the charge. Finally, we propose an ancilla wavefunction that captures both the $\mathbb P=\frac{1}{2}$ Mott insulator and the Mott semimetal* phase. The same framework has been applied by two of us to describe Mott state in twisted bilayer graphene~\cite{zhao2025ancilla}, hence the 1D model may provides support of the approach as a general framework for Mott physics in topologcal bands.

\emph{Model.}---We consider a system of spin-$\frac{1}{2}$ fermions on a one dimensional lattice with two sublattices:
\begin{equation}
\begin{split}
H=&H_0+H_{\mathrm{int}},\\
    H_0= &\gamma \sum_{i,\sigma} \left( f^\dagger_{B;i;\sigma} f_{A;i+\frac{1}{2};\sigma} - f^\dagger_{A;i-\frac{1}{2};\sigma} f_{B;i;\sigma} + \text{H.c.} \right)\\
    &+ \sum_{i,\sigma} \left(t_A f^\dagger_{A;i-\frac{1}{2};\sigma} f_{A;i+\frac{1}{2};\sigma}+ t_Bf^\dagger_{B;i;\sigma} f_{B;i+1;\sigma} + \text{H.c.} \right)\\
    &+ \Delta \sum_i (n_{B;i} - n_{A;i+\frac{1}{2}}),\\
    H_{\rm int} =& \sum_i \left( \frac{U_{A;0}}{2} \left(n_{A;i+\frac{1}{2}}-1\right)^2 + \frac{U_{B;0}}{2} \left(n_{B;i}-1\right)^2\right)\\
    & + \sum_i \left(J_{A;0}\mathbf{S}_{A;i-\frac{1}{2}} \cdot \mathbf{S}_{A;i+\frac{1}{2}} + J_{B;0}\mathbf{S}_{B;i} \cdot \mathbf{S}_{B;i+1} \right),
\end{split}
\end{equation}
where $f_{A;i}$ is $p$-wave on sublattice A with index $i+\frac{1}{2}$. $f_{B;i}$ is $s$-wave and on site $i$. $\gamma$ denotes the nearest-neighbor (NN) hopping between sublattices $A$ and $B$. $t_A$ and $t_B$ represent the nextnearest-neighbor (NNN) hopping amplitudes within sublattices $A$ and $B$ respectively. $\Delta$ is the sublattice potential. The system has inversion symmetry $P$: $f_{A;i+\frac{1}{2};\sigma}\rightarrow -f_{A;-i-\frac{1}{2};\sigma},f_{B;i;\sigma}\rightarrow f_{B;-i;\sigma}$.  When decreasing $\Delta$, there is a band inversion at $k=0$ and the lower band becomes topological in the sense that its Wannier center shifts to B sublattice from A sublattice, despite that most of the density is still on A sublattice.

\begin{figure}[t]
\includegraphics[width=\columnwidth]{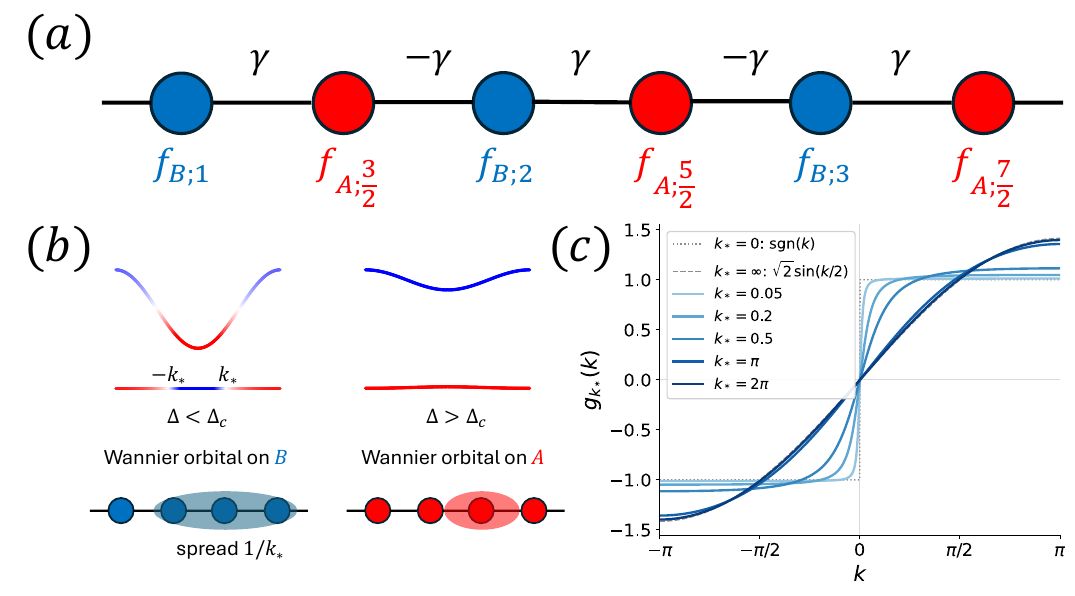}
\caption{\label{fig:illustration}
Schematic illustration of the one-dimensional two-orbital model and its active-band Wannier orbital.
(a) $A$ sublattice and $B$ sublattice are shown as red and blue circles respectively.
(b) Representative lower-band structures and orbital characters on the two different Wannier orbitals. For $\Delta<\Delta_c$, the lower band has a $B$-centered Wannier orbital with a characteristic spread $\sim 1/k_*$. For $\Delta>\Delta_c$, the lower band has an $A$-centered Wannier orbital. (c) Form factor $g(k)$ of the projected fermion,
defined through $\tilde{f}_{A;k;\sigma}=-\mathrm{i}g_{k_*}(k)c_{k;\sigma}$.
In the large-$k_*$ limit, $g_{k_*}(k)=\sqrt{2}\sin\frac{k}{2}$.}
\end{figure}

We focus on the case that the lower band is perfectly flat when $t_A=0$ and $\Delta=\frac{\gamma^2}{2t_B}-t_B$. The lower band Bloch wavefunction can be written as $c_{k;\sigma}^\dagger=-\frac{2\mathrm{i}\sin\frac{k}{2}}{\sqrt{k_*^2+4\sin^2\frac{k}{2}}}f^\dagger_{A;k;\sigma}-\frac{k_*}{\sqrt{k_*^2+4\sin^2\frac{k}{2}}}f^\dagger_{B;k;\sigma}$. The $A$-orbital weight vanishes near $k=0$ as $\frac{k^2}{k^2_*}$, but is at order one for $|k|>k_*$. We assume that the Hubbard $U_A$ is much smaller than that of the band gap and then we just project the interaction to the flat band, in the same spirit as the projection into lowest landau level in the study of quantum Hall systems.  The final model is a one-orbital Hubbard model:
\begin{equation}
\begin{split} 
H_\mathrm{eff} =& -t \sum_{i,\sigma} \left( c^\dagger_{i;\sigma} c_{i+1;\sigma} + \text{H.c.} \right) \\
    &+ \sum_i \left(\frac{U_A}{2}\left(\tilde{n}_{A;i+\frac{1}{2}}-1\right)^2+\frac{U}{2}\left(n_{i}-1\right)^2 \right)\\
    &+\sum_i \left(\frac{J_A}{2}\{\tilde{\mathbf{S}}_{A;i-\frac{1}{2}},\tilde{\mathbf S}_{A;i+\frac{1}{2}} \}+J\mathbf{S}_{i}\cdot \mathbf{S}_{i+1}\right),
    \label{eq:H}
\end{split}
\end{equation}
where $t=0$ in the flat active band limit and the model has PH symmetry $\mathcal C: c_{i;\sigma} \rightarrow \mathcal{K}c^\dagger_{i;\bar{\sigma}}$, where $\mathcal{K}$ denotes complex conjugation. $\{AB\}=AB+BA$ is the anticommutator for symmetrizing the spin interaction. The projected density and spin operators are defined by $\tilde{n}_{A;i+\frac{1}{2}}=\sum_\sigma \tilde f^\dagger_{A;i+\frac{1}{2};\sigma}\tilde f_{A;i+\frac{1}{2};\sigma}$, $\tilde{\mathbf{S}}_{A;i+\frac{1}{2}}=\sum_{\alpha\beta}\frac{1}{2}\tilde{f}^\dagger_{A;i+\frac{1}{2};\alpha}\boldsymbol{\sigma}_{\alpha\beta} \tilde{f}_{A;i+\frac{1}{2};\beta}$, where $\tilde f_A$ is $f_A$ projected on the active band and normalized as:
\begin{equation}
    \tilde f_{A;k;\sigma} = \frac{P_{\mathrm{active}} f_{A;k;\sigma} P_{\mathrm{active}}}{\sqrt{\bar{z}}}=-\mathrm{i}g_{k_*}( k) c_{k;\sigma},
\end{equation}
where $P_\mathrm{active}$ represents the projector to the active band, $\bar{z}=\frac{1}{L}\sum_k\frac{4\sin^2\frac{k}{2}}{k_*^2+4\sin^2\frac{k}{2}}$ to ensure that $\{\tilde{f}_{A;i;\sigma},\tilde{f}^\dagger_{A;i;\sigma}\}=1$. We define $g_{k_*}(k)=\frac{2\sin\frac{k}{2}}{\sqrt{(k_*^2+4\sin^2\frac{k}{2})\bar{z}}}$. The projected couplings are $U_A=\bar{z}^2 U_{A;0}, J_A=\bar{z}^2 J_{A;0}$. We symmetrize the interaction under PH transformation at $t=0$. $\tilde f_{A;i+1/2}$ is a non-local operator in terms of the canonical Wannier orbitals $c_{i;\sigma}$, which is the Fourier transformation of $c_{k;\sigma}$.  We note that the model is fully decided by the function $g_{k_*}(k)$, which is controlled solely by $k_*$ and shown in Fig.~\ref{fig:illustration}(c). At small $k_*$, $|g_{k_*}(k)|$ saturates for  $|k|>k_*$ and vanishes at $k=0$. When $k\gtrsim \pi $, $g_{k_*}$ converges to $g(k)=\sqrt{2} \sin \frac{k}{2}$. In the following, we refer to the large $k_*$ limit as using $g(k)=\sqrt{2} \sin \frac{k}{2}$.  In this limit, we have $\tilde{f}_{A;i+\frac{1}{2};\sigma}=\frac{1}{\sqrt{2}}(c_{i;\sigma}-c_{i+1;\sigma})$, thus the $U_A$ terms is really a Hubbard interaction for the bond density.  $U, J$ are just standard Hubbard and Heisenberg terms.  In the leading order, we only keep $U_A$ and always set $U_A=1$.  We will also add $J_A$ to reach a spin-singlet phase.

\begin{figure}[t]
\includegraphics[width=\columnwidth]{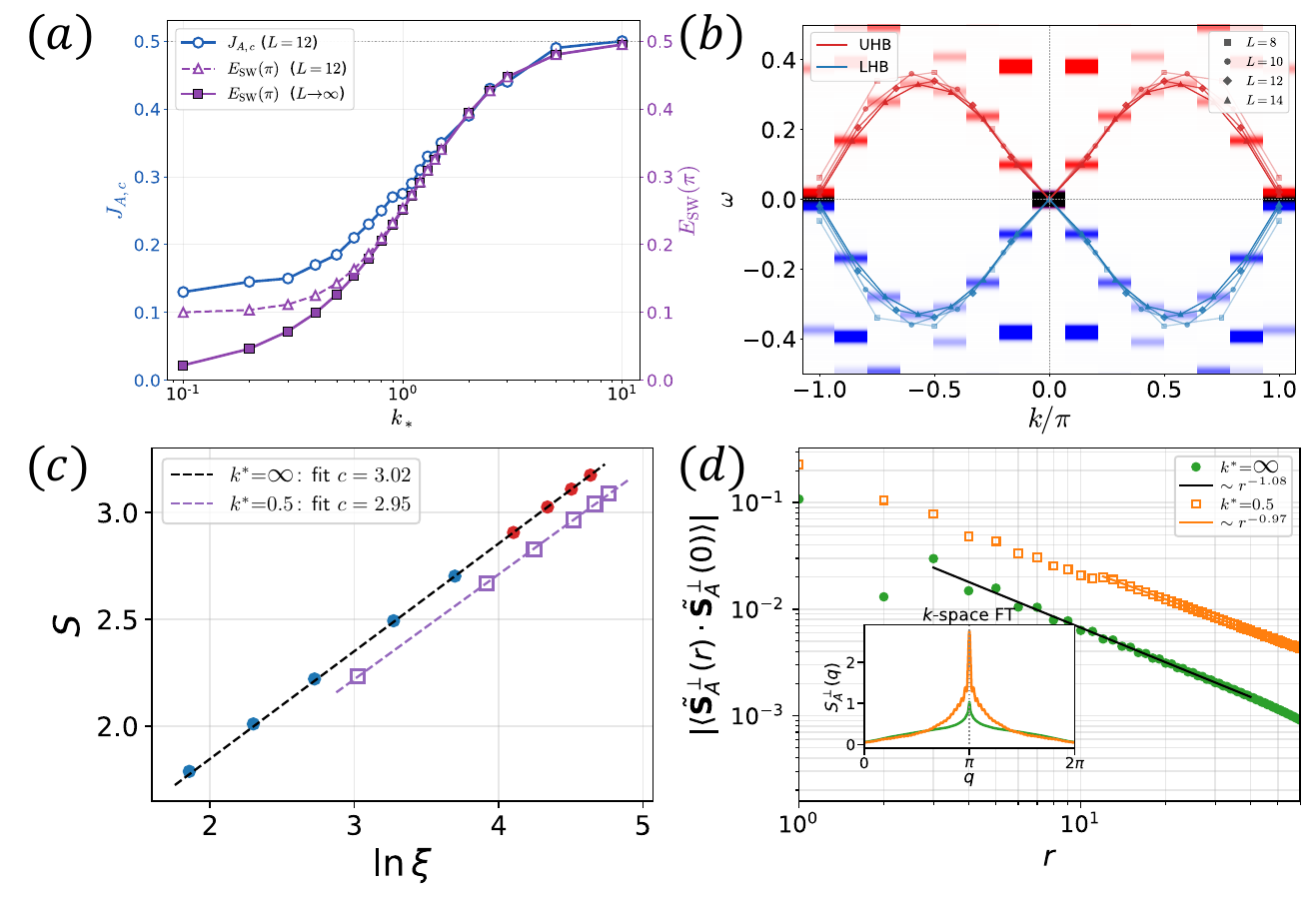}
\caption{\label{fig:cc}
(a) Critical antiferromagnetic exchange $J_{A,c}$ for melting the ferromagnetic ground state as a function of $k_*$. The spin-wave energy at momentum $\pi$, $E_{\mathrm{SW}}(\pi)$ is shown for comparison. The close agreement indicates that the critical $J_A$ is related to the spin wave bandwidth. When $L=12$, the critical $J_{A,c}$ saturate to $0.5$ when $k_*\rightarrow\infty$. (b) ED hubbard bands at $t=0$, $U_A=1$, $J_A=\frac{4}{3}$, $k_*=\infty$.
(c)(d) infinite DMRG (iDMRG) calculations at the representative point $U=\frac{1}{15},U_{A}=1, J=0,J_A=\frac{4}{3}$. (c) Entanglement entropy verus $\ln\xi$. For $k_*=\infty$, we choose $U_A=1,J_A=\frac{4}{3}$ with $\chi$ up to
$5000$, the last-four-point fit gives $c\approx3$. For $k_*=0.5$, we choose $U_A=1,J_A=0.5$ with $\chi$ up to $6400$, also gives $c\approx3$. (d) The same parameter as (c): The spin correlator $\langle\tilde{\mathbf{S}}_A(r)\cdot\tilde{\mathbf{S}}_A(0)\rangle$,
approximately has power law $\sim r^{-1}$ for both $k_*=0.5,\infty$, indicating the existence of local moments. The inset represents the structure factor $S_A(q)$ is the fourier transform of $\langle\tilde{\mathbf{S}}_A(r)\cdot\tilde{\mathbf{S}}_A(0)\rangle$, there is a peak at $q=\pi $. }
\end{figure}
\emph{Quantum Mott semimetal* phase}--- The ground state with only $U_A$ is still a ferromagnetic state, but the FM order is weak at small $k_*$ as shown in Fig.~\ref{fig:cc}(a). Next we  add a finite $J_A$ to quantum melting the FM.  In the spin singlet phase with $J_A>J_{A,c}$, we find a quantum Mott semimetal* phase, which hosts one charge mode and two spin modes and can be labeled as C1S2. As shown in Figs.~\ref{fig:cc}(b), our ED calculations show that the upper and lower Hubbard bands touch at a Dirac node for $k_*=\infty$. Moreover, this phase is stable to a finite $U, J$ as perturbation.

As a representative example, we take two example points $J_A=\frac{4}{3},U=\frac{1}{15},k_*=\infty$ and $J_A=0.5, U=0,k_*=0.5$. As shown in Fig.~\ref{fig:cc}(c), finite-entanglement scaling of the iDMRG ground state gives a central charge $c\approx 3$. In addition, the spin correlator $|\langle\mathbf{S}_A(r)\cdot\mathbf{S}_A(0)\rangle|$ decays approximately as a power law, close to $r^{-1}$, as shown in Fig.~\ref{fig:cc}(c), signaling the presence of gapless local-moment fluctuations. Meanwhile, the single-electron excitation remains gapless and exhibits a linear dispersion near $k=0$, given by the ED result in Fig.~\ref{fig:cc}(b), corresponding to an additional Dirac-like low-energy fermion. Taken together, these results identify the ground state as a C1S2 phase, with a spinful Dirac fermion at $k=0$ coexisting with a neutral spin mode from local moments. Later we will provide a low energy effective theory for the phase and show that the Dirac fermion is protected by the inversion and PH symmetry.

\begin{figure}[t]
\includegraphics[width=\columnwidth]{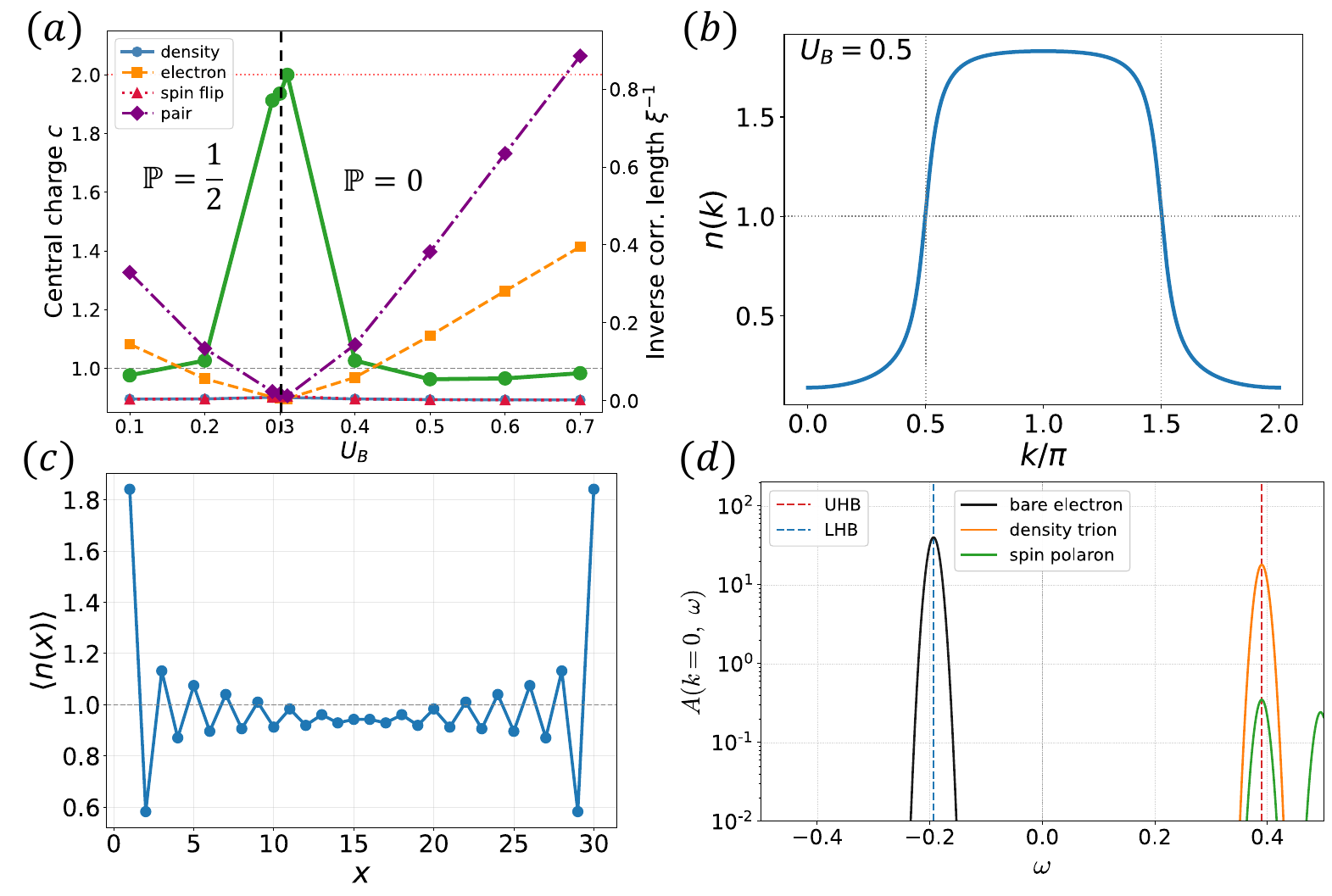}
\caption{(a) Phase diagram as a function of $U$ at fixed $t=-0.3,U_A=1, J_A=0,J=U,\chi=1000$:
central charge $c$ (green, left axis), extracted from the
finite-entanglement scaling $S=\frac{c}{6}\ln\xi$, together with the inverse
correlation lengths $\xi^{-1}$ (right axis) of the density, single-electron,
spin-flip, and pair channels of the transfer matrix. 
On both sides of the transition the system is a Mott insulator with gapless
density and spin-flip modes ($\xi^{-1}\!\approx\!0$) but gapped electron and
pair excitations, giving $c=1$.
At $U\approx0.3$ the electron and pair gaps close simultaneously and
$c\to2$, marking a direct continuous transition between a topological Mott
insulator with quantized polarization $\mathbb{P}=\tfrac12$ and a trivial
Mott insulator with $\mathbb{P}=0$.
(b)~Momentum distribution $n(k)$ at the critical point in (b) with $U=0.3$, indicating that the critical point is described by a Luttinger liquid.
(c)~Ground state density profile $\langle n(x)\rangle$ of the topological Mott
insulator at $U=J=0$ in finite DMRG calculations with $L=30$ and $\chi=2000$ under OBC.
There is Friedel oscillation near the boundary. (d) The same parameter with Fig.~\ref{fig:cc} (b) but with a small $t=0.08$, which gaps out the Dirac point. The dispersion of the lower Hubbard band agrees well with the single particle spectral function, whereas the dispersion of the upper Hubbard band is well matched with the density trion and spin polaron spectral functions.}
    \label{fig:Mott_ancilla}
\end{figure}

\emph{Symmetric Mott insulators}---A finite hopping $t\neq 0$ breaks the PH symmetry and gaps out the Dirac crossing. Remarkably, we find two distinct Mott insulating phases, distinguished by the many-body polarization defined through the Resta formula~\cite{PhysRevLett.80.1800,PhysRevB.108.235150},
\begin{equation}
    \mathbb{P}=\frac{1}{2\pi}\arg \langle \Psi|e^{\frac{2\pi\mathrm{i}}{N}\sum_j x_jn_j}|\Psi\rangle.
\end{equation}
In the presence of inversion symmetry, $\mathbb{P}$ is quantized to $0$ or $\frac{1}{2}$ and therefore provides a many-body symmetry-protected topological index. As shown in Fig.~\ref{fig:Mott_ancilla}(a), we fix $t=-0.3$, $U_A=1$, and $J_A=0$, and tune $U$ with $J=U$. The system undergoes a transition from a phase with $\mathbb{P}=\frac{1}{2}$ to one with $\mathbb{P}=0$ near $U\approx0.3$. On both sides of the transition, the charge sector is gapped while the spin sector remains gapless; at the transition, the charge gap closes. This behavior is corroborated by the iDMRG correlation lengths in different $[N,S_z]$ sectors, shown in Fig.~\ref{fig:Mott_ancilla}(a).

The two Mott insulators can be understood as arising from distinct umklapp instabilities of an underlying Luttinger liquid (the C1S1 phase), shown in Fig.~\ref{fig:Mott_ancilla}(b). Both $U_A$ and $U$ renormalize the charge Luttinger parameter toward $K_c<1$, but they generate the charge umklapp term $\cos(2\sqrt{2}\phi_c)$ with opposite signs. Consequently, the charge field $\phi_c$ is pinned at inequivalent sets of minima in the two phases. A detailed bosonization analysis is presented in Appendix~\ref{app:bosonization}.

The Mott insulator with $\mathbb{P}=\frac{1}{2}$ can be viewed as a one-dimensional analog of the symmetric `topological' Mott insulator. In references it is also called `Haldane insulator'~\cite{PhysRevLett.97.260401,PhysRevB.88.035109,PhysRevB.95.245108}. A characteristic consequence of its distinct charge-field pinning is the presence of pronounced Friedel oscillations near open boundaries. As shown in Fig.~\ref{fig:Mott_ancilla}(c), the real-space density profile $\langle n(x)\rangle$ under open boundary conditions exhibits persistent boundary-induced charge oscillations despite the bulk charge gap.

\begin{figure}[t]
\includegraphics[width=\columnwidth]{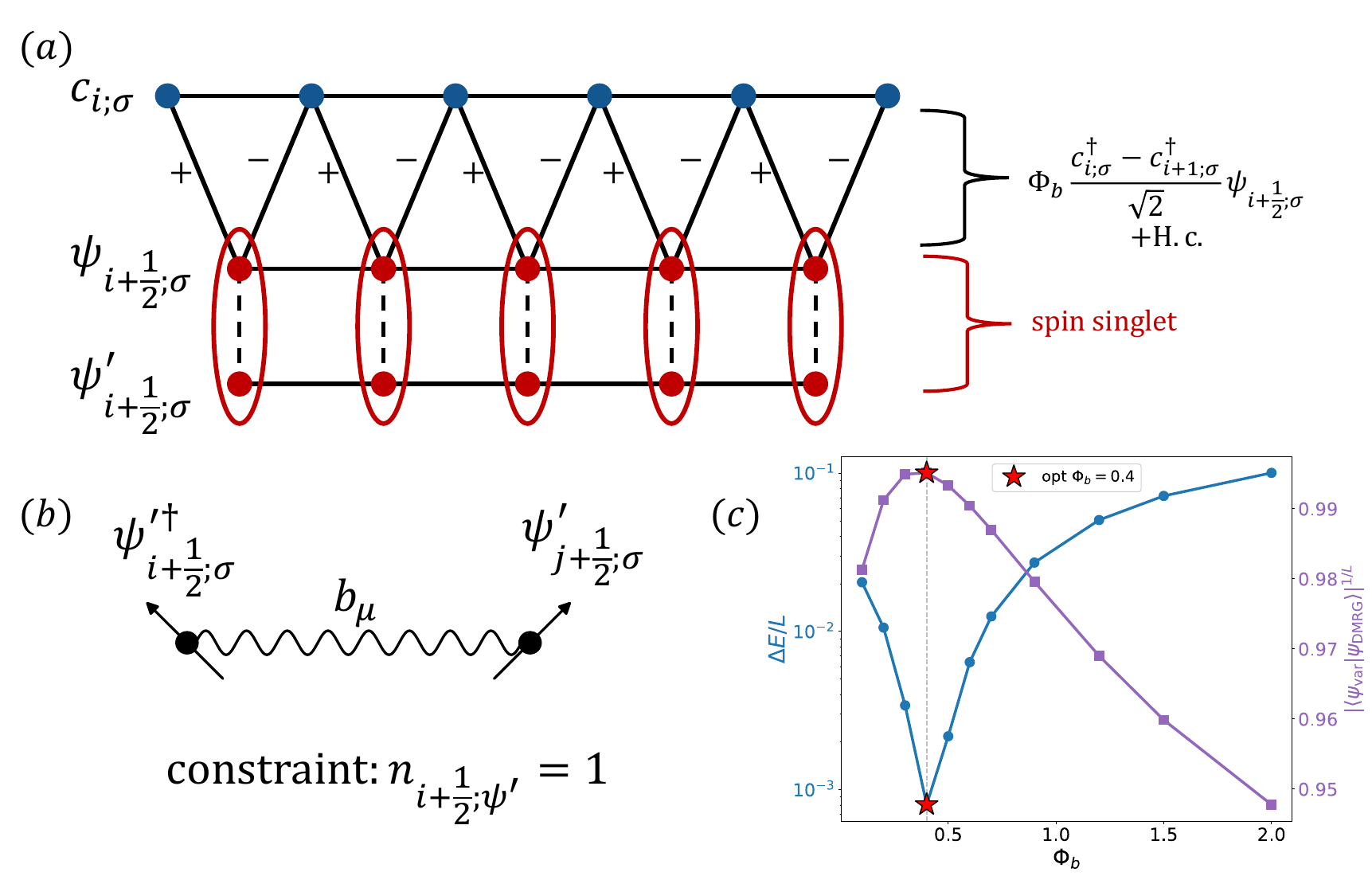}
\caption{(a)(b) The illustration of the ancilla wavefunctions. In (a), we show that physical electron (blue) resides at site $i$, while the two spinful ancilla fermions $\psi$ and $\psi^\prime$ reside at site $i+\frac{1}{2}$. Entanglement between $c$ and $\psi$ is introduced solely by the hybridization $\Phi_b$ term, whereas the entanglement between $\psi$ and $\psi^\prime$ is imposed by the
singlet projection, which requires two ancilla fermions on the same
lattice site to form a spin singlet. In (b), the $\psi^\prime$ fermions can
be viewed as Abrikosov fermions (spinons) coupled an emergent
$SU(2)$ gauge field (the wavy lines). (c) Benchmark of the single-parameter ancilla wavefunction
$|\psi_{\rm var}(\Phi_b)\rangle$ against DMRG at $L=30$ under PBC:
energy error per site $\Delta E/L$ (blue, left, log scale) and per-site
overlap $|\langle\psi_{\rm var}|\psi_{\rm DMRG}\rangle|^{1/L}$ (purple,
right) versus $\Phi_b$. Both criteria select the same optimum
$\Phi_b=0.4$ marked by red star, where $\Delta E/L\approx8\times10^{-4}$ and the
per site overlap reaches $0.995$.}
    \label{fig:pure_ancilla}
\end{figure}

\emph{Ancilla wavefunction approach}--- The $\mathbb P=\frac{1}{2}$ Mott insulator continuously evolves into the Mott semimetal* phase when we decrease the hopping $t$. At the $t\rightarrow 0$ limit with PH symmetry,  we have a Dirac cone at $k=0$ for each spin. This is quite remarkable since that a Dirac fermion needs a two-component structure and we only have one orbital for single electron within each spin.  The only way to get a Dirac cone is to have an emergent fermion $\psi_{k;\sigma}$ which hybridizes with the single electron $c_{k;\sigma}$ close to $k=0$ to form the Dirac fermion together.  As shown in Fig.~\ref{fig:Mott_ancilla}(d), when a small gap is opened under a PH breaking term, we can split the $c$ and $\psi$ components of the Dirac cone at $k=0$. The single particle spectral function $A(k,\omega)$ shows a vanishing spectral weight near $k=0$ in the upper Hubbard band. This indicates that the corresponding low energy excitation has zero overlap with the bare electron operator $c$. Then this low energy emergent fermion $\psi_\sigma$ must be a composite excitation, for example, a composite fermion formed by a single electron bound to a particle-hole operator such as density or spin. We consider the two simplest possibilities, a density trion and a spin polaron, whose explicit forms are given in the End Matter. Both of their spectral functions matches well with the upper Hubbard band, as shown in Fig.~\ref{fig:Mott_ancilla}(d).

To capture this composite excitation, next we provide a variational wavefunction based on the ancilla framework. In this approach, two auxillary fermions are introduced in addition to the physical electron $c$. The physical electrons $c$ and the $\psi$ fermion together form the charge sector, while $\psi'$ carries the charge-neutral spin degrees of freedom. The ancilla wavefunction takes the form
\begin{equation}
    \lvert \Psi_\mathrm{ancilla}\rangle=P_S\lvert\mathrm{Slater}[c,\psi]\rangle\otimes \lvert\Psi_{\psi^\prime}\rangle,
\end{equation}
where $P_S$ projects the two ancilla fermions into a $SU(2)$ spin singlet at each site $i$. 

The ancilla construction for the $\mathbb{P}=0$ Mott insulators has been studied previously by two of us~\cite{8knn-mr5x}. In that case, both ancilla fermions are at the same site as the physical electron $c$. Here we show that the $\mathbb{P}=\frac{1}{2}$ Mott insulator admits an analogous ancilla description, but with the ancilla fermions on the A sublattice (on the bond of the physical sites) and carrying the opposite inversion parity $\mathcal{I}$: $\psi_{i+\frac{1}{2}}\rightarrow-\psi_{-i-\frac{1}{2}},\psi^\prime_{i+\frac{1}{2}}\rightarrow-\psi^\prime_{-i-\frac{1}{2}}$. This modified symmetry assignment is the key ingredient that allows the projected wavefunction to realize the nontrivial many body polarization $\mathbb{P}=\frac{1}{2}$. We choose $\lvert\Psi_{\psi^\prime}\rangle$ to be the ground state of 1D Heisenberg model. The Slater determinant of charge sectors are determined by the following Hamiltonian: 
\begin{equation}
    \begin{split}
    H_\mathrm{ancilla}=&-t_c\sum_{\braket{ij},\sigma}\left(c^\dagger_{i;\sigma} c_{j;\sigma}+\mathrm{H.c.}\right)-\mu N_c-\mu_\psi N_\psi\\
    &+\Phi_b\sum_{i,\sigma}\left(\frac{c^\dagger_{i;\sigma}-c^\dagger_{i+1;\sigma}}{\sqrt{2}}\psi_{i+\frac{1}{2};\sigma}+\mathrm{H.c.}\right),\\
     \end{split}
\end{equation}
in which the chemical potential $\mu_\psi$ are chosen to fix $\langle n_\psi\rangle=1$ per site. We fix $t_c$ to be the same as $t$ in the original Hamiltonian Eq.~\ref{eq:H}, leaving $\Phi_b$ as the only variational parameter. The resulting state achieves an energy close to the DMRG ground state energy and a large overlap per site with the DMRG ground state as shown in Fig.~\ref{fig:Mott_ancilla}(d), demonstrating that the ancilla wavefunction provides an accurate description of the $\mathbb{P}=\frac{1}{2}$ Mott insulator.

Based on this ancilla framework, we can provide a low energy effective theory for the Mott semimetal* phase and its descandants from small PH breaking term as:

\begin{align}
    \mathcal L=\mathcal L_{\mathrm{charge}}+\mathcal L_{\mathrm{spin}}
\end{align}
where $\mathcal L_{\mathrm{spin}}$ represents the SU(2)$_1$ CFT for the neutral spin mode from $\psi'$.  $\mathcal{L}_{\mathrm{charge}}$ represents a C1S1 phase for the itinerant sector with a spinful Dirac fermion, captured by an effective low energy Hamiltonian as:

\begin{equation}
    \mathcal H_{\mathrm{charge}}=\sum_{k,\sigma}
    \begin{pmatrix}
        c^\dagger_{k;\sigma} & \psi^\dagger_{k;\sigma}
    \end{pmatrix} (\upsilon k \tau^y + m \tau^z) \begin{pmatrix} c_{k;\sigma} \\ \psi_{k;\sigma}\end{pmatrix},
\end{equation}
where $\tau^{x,y,z}$ are Pauli matrices in the subspace spanned by $c,\psi$. In this infra-red (IR) theory, inversion acts as :  $c_{k;\sigma}\rightarrow c_{-k;\sigma},\psi_{k;\sigma}\rightarrow-\psi_{-k;\sigma}$, PH acts as $c_{k;\sigma}\rightarrow \mathcal{K} c^\dagger_{k;\bar{\sigma}},\psi_{k;\sigma}\rightarrow -\mathcal{K} \psi^\dagger_{k;\bar{\sigma}}$, time reversal act as $c_{k;\sigma}\rightarrow \mathcal{K} c_{-k;\bar{\sigma}},\psi_{k;\sigma}\rightarrow \mathcal{K} \psi_{-k;\bar{\sigma}}$. It is clear that the inversion symmetry guarantees that the off diagonal term is linear to $k$ and time reversal symmetry forbids $k\tau^x$. PH symmetry forbids the Dirac mass term $m\tau^z$. The above effective theory can be derived from our ancilla framework.  But regardless of the specific theoretical derivation, from general argument, we must include an emergent charged fermion $\psi$ and a local moment in the low energy theory in addition to the single electron $c$ to capture the C1S2 phase. We discuss the corresponding physical operator of $\psi$ in the End Matter.  In reality, we must also include interactions in the low energy theory and in 1D it is natural to discuss the C1S2 theory using bosonization, but starting from the above low energy theory instead of the microscopic model. In the supplementary, we show that the C1S2 phase has instabilities towards symmetry-breaking, C0S2, C1S1 and C1S0 phases depending on specific interactions. But from our DMRG simulations, the C1S2 phase appears to be stable within the numerical precision we can reach for the parameter studied.

\emph{Conclusion.}---In summary, we have demosntrated a quantum Mott semimetal* phase in a one-dimensional Hubbard model with only one orbital. Starting from a flat-band model with momentum dependent orbital form factors, we find that the Hubbard interaction alone favors a ferromagnetic ground state.  But sufficiently strong antiferromagnetic spin interaction drives a quantum melting of the ferromagnet and stabilizes a symmetric gapless ground state with a spinful Dirac fermion at $k=0$ coexisting with a gapless neutral spin mode from the local moments.
Similar phases have been discussed in TBG with mechanism related to the concentrated Berry curvature. Our model at large $k_*$ regime clearly shows that either the Berry curvature or a small momentum space scale $k_*$ is not essential for the physics.  Instead, we have a general framework to unify the Mott states in trivial and topological bands in an effective theory where we have an emergent heavy fermion model from a one-orbital model: local moments coexist with an itinerant sector, which is formed together by  the electron  $c_\sigma$ and an emergent fermion field $\psi_\sigma$. The hybridization $\Phi(k) c^\dagger(k) \psi(k)$ provides the Mott gap. This $\psi_\sigma$ is constrained to be a well-localized orbital, therefore when the flat band has a different quantum number at a momentum $k_0$, the hybridization and thus the Mott gap vanishes at $k_0$.  We interpret $\psi_\sigma$ as a composite operator with electron bound to density and spin operators. Similar framework has also been used to describe the FL* phase for the hole doped cuprate\cite{zhang2020pseudogap}. Therefore the emergence of heavy Fermion picture and composite fermion may be a universal physics behind Mott states in correlated electrons.

\begin{acknowledgments}
We thank Jing-Yu Zhao for previous collaborations and discussions. The work is supported by the Alfred P. Sloan Foundation through a Sloan Research
Fellowship (Y.-H.Z.). T.W. also acknowledges support from the Harvard Quantum Initiative Fellowship.
\end{acknowledgments}
\bibliography{references}

\bigskip
\bigskip
\begin{center}
\textbf{\large End Matter}
\end{center}
\bigskip
\vspace{-0.2cm}
\emph{Origin of $\psi$}---In the main text, we introduced an ancilla fermion $\psi$ in the low energy effective theory of the Mott semimetal$^\ast$. It carries an inversion quantum number opposite to that of $c$. A general microscopic form is
\begin{equation}
    \psi_{k;\sigma}
    =
    A_k c_{k;\sigma}
    +
    B_k\sum_{q,\sigma'}
    c_{k-q;\sigma'}O_{k;q;\sigma\sigma'}
    +\cdots,
\end{equation}
where $O_{k;q;\sigma\sigma'}$ is a particle-hole operator, such as a density or spin operator.

At $k=0$, $A_0$ vanishes because
$\mathcal I:c_{0;\sigma}\rightarrow c_{0;\sigma}$,
whereas $\psi_{0;\sigma}$ is inversion odd. The simplest form of
$\psi_{0;\sigma}$ can be constructed in the spirit of the three-particle
operators introduced in Hubbard's original work~\cite{10.1098/rspa.1963.0204},
where
$\psi_{i;\sigma}=U^{-1}[c_{i;\sigma},H_{\mathrm{int}}]$.
In our case, the leading three-particle operator is
$\psi_{i+\frac12;\sigma}\sim
[\tilde f_{A;i+\frac12;\sigma},H_{\mathrm{int}}]$.
For the Hubbard and Heisenberg interactions involving $\tilde f_A$, we obtain
\begin{equation}
\begin{aligned}
\bigl[\tilde f_{A;i+\frac12;\alpha},H_{U_A}\bigr]
&=
U_A \tilde f_{A;i+\frac12;\alpha}
\delta\tilde n_{A;i+\frac12}
\\
\bigl[\tilde f_{A;i+\frac12;\alpha},H_{J_A}\bigr]
&=
J_A \tilde f_{A;i+\frac12;\beta}
\boldsymbol{\sigma}_{\beta\alpha}
\\
&\cdot
\left(
\tilde{\mathbf S}_{A;i-\frac12}
+
\tilde{\mathbf S}_{A;i+\frac32}
\right).\\
\end{aligned}
\label{eq:EOM_A}
\end{equation}
These two terms can be understood as density trion and spin polaron operators respectively. The density trion and spin polaron operators in the main text are defined as:
\begin{equation}
    \begin{split}
      \psi^U_{i+\frac{1}{2};\alpha}=&\left(c_{i;\alpha}-c_{i+1;\alpha}\right)\delta\tilde{n}_{A;+\frac{1}{2}},\\  
      \psi^J_{i+\frac{1}{2};\sigma}=&\left(c_{i;\beta}-c_{i+1;\beta}\right)\boldsymbol{\sigma}_{\beta\alpha}\left(
\tilde{\mathbf S}_{A;i-\frac12}
+
\tilde{\mathbf S}_{A;i+\frac32}
\right).
    \end{split}
\end{equation}

\onecolumngrid
\appendix

\begin{figure}[t]
\includegraphics[width=\columnwidth]{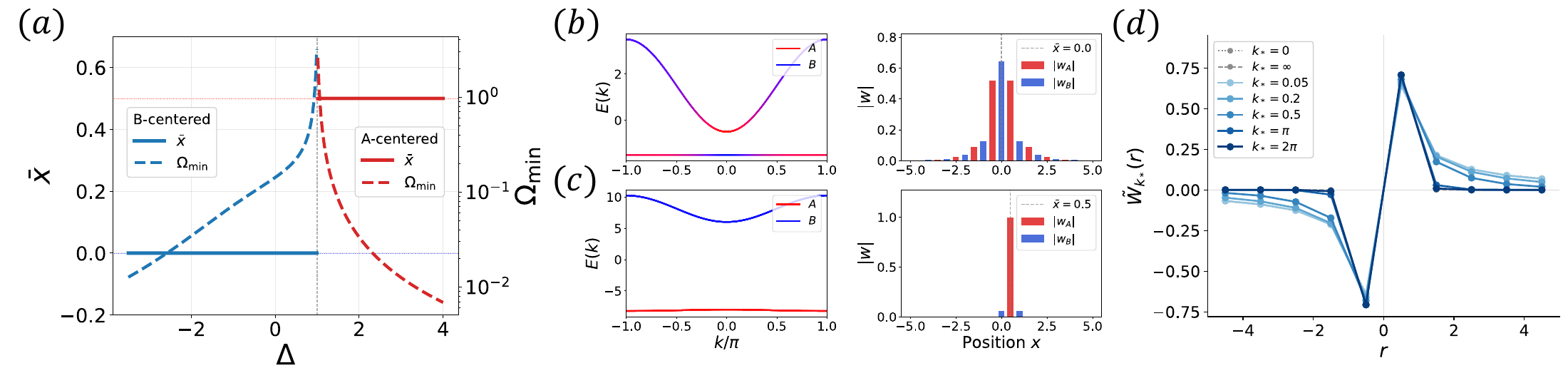}
\caption{\label{fig:Wannier}
Wannier representation of the lower band. We choose $t_A=0, t_B=-1, \gamma=1$ in the calculations.
(a) Wannier center $\bar{x}$ and minimal spread $\Omega_{\min}$ as functions of the sublattice potential $\Delta$. Inversion symmetry pins the Wannier center either to the $B$ sublattice, $\bar{x}=0$, or to the $A$ sublattice, $\bar{x}=\frac{1}{2}$. The minimal spread is enhanced near the transition between the two Wannier representations.
(b) $\Delta=0.5$. Representative band structure and real-space Wannier profile in the $B$-centered phase. The Wannier orbital has sizable weight on neighboring sites and is spatially extended.
(c) $\Delta=8$. Representative band structure and real-space Wannier profile in the $A$-centered phase. The Wannier orbital is strongly localized near the $A$ sublattice. (d) At $\Delta=0$, the projected $A$ operator wavefuntion in real space distribution $\tilde{W}_{k_*}(r)$.}
\end{figure}
\section{Wannier orbital analysis of the lower band}
\label{app:wannier}
In momentum space with basis $\Psi_{k;\sigma}=(f_{A;k;\sigma},f_{B;k;\sigma})^T$, and adopting the cell-periodic convention in which the $A$-site sits $x_A=\frac{1}{2}$, $B$-site sites at $x_B=0$ within the unit cell, the free single-particle Hamiltonian reads
\begin{equation}
H_0 =\sum_{k;\sigma} \Psi^\dagger_{k;\sigma}\begin{pmatrix} -\Delta + 2t_A\cos k & -2\mathrm{i}\gamma\sin\frac{k}{2} \\ 2\mathrm{i}\gamma\sin\frac{k}{2} & \Delta + 2t_B\cos k \end{pmatrix}\Psi_{k;\sigma}.
\label{eq:hk}
\end{equation}
The two Bloch eigenstates can be written as:
\begin{equation}
\begin{split}
    c^\dagger_{1;k;\sigma}
    =&\frac{1}{\mathcal{N}_k}\left(2\mathrm{i}\gamma\sin\frac{k}{2}f^\dagger_{A;k;\sigma}+\left(\delta_k+\sqrt{\delta_k^2+4\gamma^2\sin^2\frac{k}{2}}\right)f^\dagger_{B;k;\sigma}\right),\\
    c^\dagger_{2;k;\sigma}
    =&\frac{1}{\mathcal{N}_k}\left(\left(\delta_k+\sqrt{\delta_k^2+4\gamma^2\sin^2\frac{k}{2}}\right)f^\dagger_{A;k;\sigma}+2\mathrm{i}\gamma\sin\frac{k}{2}f^\dagger_{B;k;\sigma}\right),\\
    \mathcal{N}_k=&\sqrt{2\delta_k^2+8\gamma^2\sin^2\frac{k}{2}+2\delta_k\sqrt{\delta_k^2+4\gamma^2\sin^2\frac{k}{2}}},\\
    \delta_k=&-\Delta+(t_A-t_B)\cos k.
\end{split}
\end{equation}
The Wannier orbital can be constructed as:
\begin{equation}
\begin{split}
    c^\dagger_{1;i;\sigma}=&\sum_j \left(w_{1;A;i-j-\frac{1}{2}} f^\dagger_{A;j+\frac{1}{2};\sigma}+w_{1;B;i-j}f^\dagger_{B;j;\sigma}\right),\\
    c^\dagger_{2;i;\sigma}=&\sum_j \left(w_{2;A;i-j-\frac{1}{2}} f^\dagger_{A;j+\frac{1}{2};\sigma}+w_{2;B;i-j}f^\dagger_{B;j;\sigma}\right),\\
    w_{1;A;i-j-\frac{1}{2}}=&\sum_k \frac{1}{L} u_{1;A;k}e^{\mathrm{i}k(i-j)}, \quad w_{1;B;i-j}=\sum_k \frac{1}{L} u_{1;B;k}e^{\mathrm{i}k(i-j)},\\
    w_{2;A;i-j-\frac{1}{2}}=&\sum_k \frac{1}{L} u_{2;A;k}e^{\mathrm{i}k(i-j)}, \quad w_{2;B;i-j}=\sum_k \frac{1}{L} u_{2;B;k}e^{\mathrm{i}k(i-j)},\\
    u_{1;A;k}=&\frac{2\mathrm{i}\gamma\sin\frac{k}{2}e^{-\mathrm{i}\frac{k}{2}}}{\mathcal{N}_k}, \quad u_{1;B;k}=\frac{\left(\delta_k+\sqrt{\delta_k^2+4\gamma^2\sin^2\frac{k}{2}}\right)}{\mathcal{N}_k},\\
    u_{2;A;k}=&\frac{\left(\delta_k+\sqrt{\delta_k^2+4\gamma^2\sin^2\frac{k}{2}}\right)e^{-\mathrm{i}\frac{k}{2}}}{\mathcal{N}_k}, \quad u_{2;B;k}=\frac{2\mathrm{i}\gamma\sin\frac{k}{2}}{\mathcal{N}_k}.
\end{split}
\end{equation}
The Berry connection and the quantum metric of the bands are defined as:
\begin{equation}
\begin{split}
    A_{1,2}(k)=&\mathrm{i}\langle u_{1,2;k}\lvert\partial_k u_{1,2;k}\rangle,\\
    g_{1,2;k}=&\langle\partial_ku_{1,2;k}\lvert \partial_k u_{1,2;k}\rangle-|A_{1,2}(k)|^2,
\end{split}
\end{equation}
where $\lvert u_{1,2;k}\rangle=(u_{1,2;A;k},u_{1,2;B;k})^T$. The Wannier center and minimum Wannier spread can be calculated as:
\begin{equation}
\begin{split}
    \bar{x}_{1,2}=&\frac{1}{L}\sum_k A_{1,2}(k),\\
    \Omega_{\mathrm{min};1,2}=&\frac{1}{L}\sum_k g_{1,2}(k).
\end{split}
\end{equation}
In the main text, we focus on the case $t_A=0$. As shown in Fig.~\ref{fig:Wannier}(a), the Wannier center of the lower band changes as the sublattice potential $\Delta$ is varied. Two representative regimes are shown in Figs.~\ref{fig:Wannier}(b)(c). For $\Delta<\Delta_c$, the lower band admits a $B$-centered Wannier orbital, which is relatively broad in real space. For $\Delta>\Delta_c$, the Wannier orbital becomes $A$-centered and is much more localized. At $\Delta=\frac{\gamma^2}{2t_B}-t_B$, the lower band is perfectly flat. One can project the $A$-orbital to the active band as discussed in the main text:
\begin{equation}
    \tilde f_{A;k;\sigma} = -\frac{2\mathrm{i}\sin\frac{k}{2}}{\sqrt{k_*^2+4\sin^2\frac{k}{2}}} c_{k;\sigma},
\end{equation}
where $k_*=-\frac{\gamma}{t_B}$. Then we can have:
\begin{equation}
    \begin{split}
         \tilde f_{A;i+\frac{1}{2};\sigma} = &\sum_j \tilde{W}_{k_*}(i+\frac{1}{2}-j)c_{j;\sigma},\\
         \tilde{W}_{k_*}(i+\frac{1}{2}-j)=&\frac{1}{L}\sum_k \left(-\frac{2\mathrm{i}\gamma\sin\frac{k}{2}}{\sqrt{k_*^2+4\gamma^2\sin^2\frac{k}{2}}}e^{\mathrm{i}(i+\frac{1}{2}-j)}\right).
    \end{split}
\end{equation}
The illustration of $\tilde{W}_{k_*}(r)$ is shown in Fig.~\ref{fig:Wannier}(d).

\section{Spin wave calculation}\label{app:SW}
In the case that $t=J_{A}=J=0$, the spin wave can be solved exactly. We choose the fully polarized state as $\lvert\mathrm{FM}\rangle=\prod_i c^\dagger_{i;\downarrow}\lvert0\rangle$. This state is an exact eigenstate of the effective Hamiltonia defined in the main text, which is also written here as:
\begin{equation}
    \begin{split}
        H_\mathrm{eff}=&\sum_i \left(\frac{U_A}{2}\left(\tilde n_{A;i+\frac{1}{2}}-1\right)^2+\frac{U}{2}\left(n_{i}-1\right)^2\right)\\
    =&\frac{U_A}{L}\sum_{k,k^\prime,q}g_{k+q}g_k g_{k^\prime-q}g_{k^\prime}c^\dagger_{k+q;\uparrow}c_{k;\uparrow}c^\dagger_{k^\prime-q;\downarrow}c_{k^\prime;\downarrow}-\frac{U_A}{2}\sum_{k;\sigma}g_k^2c^\dagger_{k;\sigma}c_{k;\sigma}\\
    &+\frac{U}{L}\sum_{k,k^\prime,q}c^\dagger_{k+q;\uparrow}c_{k;\uparrow}c^\dagger_{k^\prime-q;\downarrow}c_{k^\prime;\downarrow}-\frac{U}{2}\sum_{k;\sigma}c^\dagger_{k;\sigma}c_{k;\sigma},
    \end{split}
\end{equation}
where $g_k=\frac{2\sin\frac{k}{2}}{\sqrt{(k_*^2+4\sin^2\frac{k}{2})\bar{z}}},\bar{z}=\frac{1}{L}\sum_k\frac{4\sin^2\frac{k}{2}}{k_*^2+4\sin^2\frac{k}{2}}$. We note that $g_k$ is same as $g_{k*}(k)$ in the main text. The energy of $\lvert\mathrm{FM}\rangle$ is $E_{\mathrm{FM}}=
-\frac{L}{2}(U_A+U)$ for this effective Hamiltonian. For the $\Delta S=1$ subspace, we can define the particle-hole basis $\lvert q,k\rangle\equiv c^\dagger_{k+q;\uparrow}c_{k;\downarrow}\lvert\mathrm{FM}\rangle$. We first evaluate the interaction matrix elements explicitly. For the
quartic terms, it is convenient to denote the momentum transfer by
$Q$. Using the fact that the spin-$\downarrow$ band is completely filled in
$\lvert\mathrm{FM}\rangle$, one finds
\begin{equation}
    \langle q,k^\prime\vert
    c^\dagger_{p+Q;\uparrow}c_{p;\uparrow}
    c^\dagger_{p^\prime-Q;\downarrow}c_{p^\prime;\downarrow}
    \vert q,k\rangle=
    \delta_{p,k+q}\,
    \delta_{k^\prime,k+Q}
    \left(
        \delta_{Q,0}
        -
        \delta_{p^\prime,k^\prime}
    \right).
\label{eq:ph_matrix_element}
\end{equation}
We define that $H_{U_A}=\sum_i \frac{U_A}{2}\left(\tilde n_{A;i+\frac{1}{2}}-1\right)^2$, $H_{U}=\sum_i \frac{U}{2}\left(n_{i}-1\right)^2$. For the $U_A$ interaction, Eq.~\eqref{eq:ph_matrix_element} gives
\begin{equation}
    \langle q,k^\prime\vert H_{U_A}\vert q,k\rangle    =
   U_A
    g_{k+q}^2\delta_{k^\prime,k}
    -
    \frac{U_A}{L}
    g_k g_{k+q}g_{k^\prime}g_{k^\prime+q}+\frac{U_A}{2}\left(g^2_k-g^2_{k+q}\right)\delta_{k^\prime,k}-\frac{L}{2}U_A.
\label{eq:VA_matrix}
\end{equation}
The $U$ interaction can be evaluated in the same way. Since its form
factor is momentum independent, one obtains
\begin{equation}
    \langle q,k^\prime\vert H_{U}\vert q,k\rangle  
    =
    U\delta_{k^\prime,k}
    -
    \frac{U}{L}-\frac{L}{2}U.
\label{eq:HB_ph}
\end{equation}
Considering $E_{\mathrm{FM}}$, the exact Hamiltonian in the one-spin-flip subspace is
\begin{equation}
    \mathcal{H}^{(q)}_{k^\prime k}
    \equiv
    \langle q,k^\prime\vert
    \left(H_{\mathrm{eff}}-E_{\mathrm{FM}}\right)
    \vert q,k\rangle=
    \left(
        U
        +
        \frac{U_A}{2}
        \left(
            g_k^2+g_{k+q}^2
        \right)
    \right)
    \delta_{k^\prime,k}-
    \frac{U_A}{L}
    g_k g_{k+q}g_{k^\prime}g_{k^\prime+q}
    -
    \frac{U}{L}.
\label{eq:exact_spinwave_matrix}
\end{equation}
We define the eigenstate as
\begin{equation}
    \lvert\Psi_q\rangle=\sum_k\psi_q(k)\lvert q,k\rangle,
\end{equation}
Then we can write down the Schr\"odinger equation:
\begin{equation}
\begin{split}
    \left(
        \left(U+\frac{U_A}{2}\left(g^2_k+g^2_{k+q}\right)\right)-E_{\mathrm{SW}}(q)
    \right)\psi_q(k)
    &=
    \frac{U_A}{L}
    g_k g_{k+q}
    \sum_{k^\prime}
    g_{k^\prime} g_{k^\prime+q}\psi_q(k^\prime)
     +
    \frac{U}{L}
    \sum_{k^\prime}\psi_q(k^\prime).
\end{split}
\label{eq:spinwave_schrodinger}
\end{equation}
To solve Eq.~\eqref{eq:spinwave_schrodinger}, we define
\begin{equation}
    A_q
    \equiv
    \frac{U_A}{L}
    \sum_k
    g_k g_{k+q}\psi_q(k),
    \qquad
    B_q
    \equiv
    \frac{U}{L}
    \sum_k
    \psi_q(k).
\end{equation}
The spin-wave wavefunction can then be written as
\begin{equation}
    \psi_q(k)
    =
    \frac{
       \bar{z} g_k g_{k+q}A_q+B_q
    }{
        U
        +
        \dfrac{U_A}{2}
        \left(g_k^2+g_{k+q}^2\right)
        -
        E_{\mathrm{SW}}(q)
    }.
\label{eq:spinwave_wavefunction_general}
\end{equation}
Substituting Eq.~\eqref{eq:spinwave_wavefunction_general} back into
the definitions of $A_q$ and $B_q$, we obtain
\begin{equation}
\begin{split}
    A_q
    &=
    \frac{U_A}{L}
    \sum_k
    \frac{
        g_k^2g_{k+q}^2 A_q
        +
       g_k g_{k+q} B_q
    }{
        U
        +
        \dfrac{U_A}{2}
        \left(g_k^2+g_{k+q}^2\right)
        -
        E_{\mathrm{SW}}(q)
    },
    \\
    B_q
    &=
    \frac{U}{L}
    \sum_k
    \frac{
    g_k g_{k+q} A_q
        +
        B_q
    }{
        U
        +
        \dfrac{U_A}{2}
        \left(g_k^2+g_{k+q}^2\right)
        -
        E_{\mathrm{SW}}(q)
    }.
\end{split}
\label{eq:spinwave_self_consistency}
\end{equation}
Therefore, $A_q$ and $B_q$ satisfy
\begin{equation}
\begin{pmatrix}
    1
    -
    \dfrac{U_A}{2L}
    \displaystyle\sum_k
    \dfrac{
        g_k^2g_{k+q}^2
    }{
        U
        +
        \dfrac{U_A}{2}
        \left(g_k^2+g_{k+q}^2\right)
        -
        E_{\mathrm{SW}}(q)
    }
    &
    -
    \dfrac{U_A}{L}
    \displaystyle\sum_k
    \dfrac{
       g_k g_{k+q}
    }{
        U
        +
        \dfrac{U_A}{2}
        \left(g_k^2+g_{k+q}^2\right)
        -
        E_{\mathrm{SW}}(q)
    }
    \\
    -
    \dfrac{U}{L}
    \displaystyle\sum_k
    \dfrac{
      g_k g_{k+q}
    }{
        U
        +
        \dfrac{U_A}{2}
        \left(g_k^2+g_{k+q}^2\right)
        -
        E_{\mathrm{SW}}(q)
    }
    &
    1
    -
    \dfrac{U}{L}
    \displaystyle\sum_k
    \dfrac{
        1
    }{
        U
        +
        \dfrac{U_A}{2}
        \left(g_k^2+g_{k+q}^2\right)
        -
        E_{\mathrm{SW}}(q)
    }
\end{pmatrix}
\begin{pmatrix}
    A_q\\
    B_q
\end{pmatrix}
=
0.
\label{eq:spinwave_matrix_equation}
\end{equation}
A nontrivial solution therefore requires the determinant of the above
matrix to vanish. 
In the following, we focus on the case $U=0$. In this limit, then we obtain the exact
bound-state equation
\begin{equation}
    1
    =
    \frac{U_A}{L}
    \sum_k
    \frac{
        g_k^2g_{k+q}^2
    }{
        \dfrac{U_A}{2}
        \left(g_k^2+g_{k+q}^2\right)
        -
        E_{\mathrm{SW}}(q)
    }.
\label{eq:spinwave_boundstate_UB0}
\end{equation}
One can verify that $E_{\mathrm{SW}}=0$. We now consider the zone boundary
$q=\pi$ that can reflect the magnon bandwidth. Eq.~\eqref{eq:spinwave_boundstate_UB0} can be solved
analytically in the thermodynamic limit. First, we note that
\begin{equation}
    \bar{z}
    =
    \int_{-\pi}^{\pi}\frac{dk}{2\pi}
    \frac{4\sin^2\frac{k}{2}}
    {k_*^2+4\sin^2\frac{k}{2}}
    =
    1
    -
    k_*^2
    \int_{-\pi}^{\pi}\frac{dk}{2\pi}
    \frac{1}
    {k_*^2+4\sin^2\frac{k}{2}}
    =
    1-\frac{k_*}{\sqrt{k_*^2+4}}.
\label{eq:zbar_exact}
\end{equation}
It is convenient to define
\begin{equation}
    \eta
    \equiv
    \frac{k_*}{\sqrt{k_*^2+4}},
    \qquad
    \bar{z}=1-\eta,
    \qquad 
    e_\pi\equiv\frac{E_{\mathrm{SW}}(\pi)}{U_A}.
\label{eq:eta_definition}
\end{equation}
Using
\begin{equation}
    \bar{z}^2g_k^2g_{k+\pi}^2
    =
    \frac{4\sin^2 k}
    {k_*^2(k_*^2+4)+4\sin^2 k},
    \qquad
    \bar{z}\left(g_k^2+g_{k+\pi}^2\right)
    =
    \frac{4\left(k_*^2+2\sin^2 k\right)}
    {k_*^2(k_*^2+4)+4\sin^2 k},
\end{equation}
Eq.~\eqref{eq:spinwave_boundstate_UB0} becomes
\begin{equation}
\begin{split}
    1
    =&
    \frac{2}{\bar{z}}
    \int_{-\pi}^{\pi}\frac{dk}{2\pi}
    \frac{\sin^2 k}
    {
        k_*^2\left(
            1-\frac{\bar{z}e_\pi}{2}(k_*^2+4)
        \right)
        +
        2\left(1-\bar{z}e_\pi\right)\sin^2 k
    }\\
    =&\frac{1}
    {\bar{z}\left(1-\bar{z}e_\pi\right)}
    \left(
        1-
        \sqrt{
        \frac{
            k_*^2\left(
                1-\frac{\bar{z}e_\pi}{2}(k_*^2+4)
            \right)
        }{
            k_*^2\left(
                1-\frac{\bar{z}e_\pi}{2}(k_*^2+4)
            \right)
            +
            2\left(1-\bar{z}e_\pi\right)
        }}
    \right)\\
    =& 
    \frac{1}
    {(1-\eta)\left(1-(1-\eta)e_\pi\right)}
    \left(
        1-
        \sqrt{
        \frac{
            2\eta^2\left(1+\eta-2e_\pi\right)
        }{
            (1+\eta^2)
            \left(
                1+\eta-(1+\eta^2)e_\pi
            \right)
        }}
    \right).
\end{split}
\label{eq:qpi_integral}
\end{equation}
Finally we can solve that
\begin{equation}
    E_{\mathrm{SW}}(\pi)
    = e_\pi U_A=
    \frac{
        \eta(1+\eta)
        \left(
            \sqrt{1+(1-\eta)^2}-1
        \right)
    }{
        (1+\eta^2)(1-\eta)^2
    }
    U_A.
\label{eq:spinwave_qpi_exact}
\end{equation}

\begin{table*}[t]

\centering
\small
\renewcommand{\arraystretch}{1.35}
\setlength{\tabcolsep}{6pt}

\begin{tabular}{@{}p{0.22\textwidth}p{0.74\textwidth}@{}}
\toprule
Object & Bosonized expression \\
\midrule

Lattice fermion &
$\displaystyle 
c_{i;\sigma}
\simeq 
e^{\mathrm{i} k_F x}\psi_{R;\sigma}(x)
+
e^{-\mathrm{i} k_F x}\psi_{L;\sigma}(x)
$ \\

Chiral fermion &
$\displaystyle
\psi_{r;\sigma}(x)
=
\frac{\kappa_\sigma}{\sqrt{2\pi }}
e^{-\mathrm{i}\left(r\phi_\sigma-\theta_\sigma\right)},
\quad r=1(R)/-1(L),\quad \{\kappa_\sigma,\kappa_{\sigma^\prime}\}=2\delta_{\sigma,\sigma^\prime}
$ \\
Charge and spin fields &
$\displaystyle
\phi_{c,s}
=
\frac{\phi_\uparrow\pm\phi_\downarrow}{\sqrt{2}},
\quad
\theta_{c,s}
=
\frac{\theta_\uparrow\pm\theta_\downarrow}{\sqrt{2}}
$ \\

$2k_F$ bilinear &
$\displaystyle
\psi^\dagger_{L;\sigma}\psi_{R;\sigma}
\sim
\frac{\mathrm{i}}{2\pi\alpha}e^{-2\mathrm{i}\phi_\sigma},
\quad
\psi^\dagger_{R;\sigma}\psi_{L;\sigma}
\sim
\frac{-\mathrm{i}}{2\pi\alpha}e^{2\mathrm{i}\phi_\sigma}
$ \\

Density operator &
$\displaystyle
n_\sigma(x)
\simeq
-\frac{1}{\pi}\partial_x\phi_\sigma
+
e^{2\mathrm{i}k_Fx}\psi^\dagger_{L;\sigma}\psi_{R;\sigma}
+
e^{-2\mathrm{i}k_Fx}\psi^\dagger_{R;\sigma}\psi_{L;\sigma}
$ \\

\bottomrule
\end{tabular}
\caption{Bosonization dictionary used in Appendix.~\ref{app:bosonization}.}
\label{tab:bosonization_dictionary}
\end{table*}

\section{Continuous transition between two Mott insulators}
\label{app:bosonization}
In the main text, we show that when $t\ne0$,the system can realize two different Mott insulators separated by a quantum critical point. At the critical point the system is a Luttinger liquid. The critical point is stable for $K_c\le1$, while spin $SU(2)$ symmetry fixes $K_s=1$. Here we treat $U_A$ and $U$ as two diffrent perturbations to this Luttinger liquid and show that they lead to different instabilities. We focus on the large-$k_*$ limit, in which the projected $A$-orbital fermion reduces to a nearest-neighbor bond operator. The effective Hamiltonian is
\begin{equation}
H_\mathrm{eff} = H_{\mathrm{critical}} + \sum_i \left(\frac{U_A}{2}\left(\frac{1}{2}\sum_\sigma\left(c^\dagger_{i;\sigma}-c_{i+1;\sigma}^\dagger\right)\left(c_{i;\sigma}-c_{i+1;\sigma}\right)-1\right)^2+\frac{U}{2}\left(n_{i}-1\right)^2 \right).
\end{equation}
Here we take $H_{\mathrm{critical}}=-\sum_{\langle ij\rangle,\sigma}\left(c^\dagger_{i;\sigma}c_{j;\sigma}+\mathrm{H.c.}\right)$ for brevity. Later we show that both $U_A$ and $U$ decrease $K_c$. Therefore for the stability, it is safe for us to set $K_c=1$ for $H_{\mathrm{critical}}$ would be safe. 
The essential distinction between the two interaction terms is that $U$ acts on the local site density, whereas $U_A$ acts on the density of the projected bond orbital. As we show below, the bond form factor changes the phase of the $2k_F$ density component and consequently reverses the sign of the charge umklapp term. The two repulsive interactions therefore favor inequivalent strong-coupling instabilities.

For the $U_A$ interaction terms, when there is only free term, the operators can be bosonized as listed in Table.~\ref{tab:bosonization_dictionary}. The projected fermion defined in the main text can be expressed as
\begin{equation}
\begin{split}
   \tilde{f}_{A;i+\frac{1}{2};\sigma}\lvert_{k_*\rightarrow\infty} =&\frac{1}{\sqrt{2}}\left( c_{i;\sigma}-c_{i+1;\sigma}\right)\simeq\frac{1}{\sqrt{2}}\left(\psi_{R;\sigma}e^{\mathrm{i}k_Fx}(1-e^{\mathrm{i}k_F})+\psi_{L;\sigma}e^{-\mathrm{i}k_Fx}(1-e^{-\mathrm{i}k_F})\right),\\
   \tilde{n}_{A;i+\frac{1}{2};\sigma}=&\tilde{f}^\dagger_{A;\sigma}\tilde{f}_{A;\sigma}\simeq\mathrm{i}\left(\psi^\dagger_{R;\sigma}\psi_{L;\sigma}e^{-2\mathrm{i}k_Fx}-\psi^\dagger_{L;\sigma}\psi_{R;\sigma}e^{2\mathrm{i}k_Fx}\right)-\frac{1}{\pi}\partial_x\phi_\sigma=\frac{1}{\pi\alpha}\cos\left(2\phi_\sigma-2k_Fx\right)-\frac{1}{\pi}\partial_x\phi_\sigma.
\end{split}
\end{equation}
The Hubbard interaction for $\tilde{f}_A$ can be written as
\begin{equation}
\tilde{n}_{A;i+\frac{1}{2};\uparrow}\tilde{n}_{A;i+\frac{1}{2};\downarrow}=
\frac{1}{2\pi^2}\left(\left(\partial_x\phi_c\right)^2-\left(\partial_x\phi_s\right)^2\right)+\frac{1}{2\pi^2\alpha^2}\left(\cos\left(2\sqrt{2}\phi_s\right)+\cos\left(2\sqrt{2}\phi_c-4k_Fx\right)\right).
\end{equation}
At half filling, $k_F=\frac{\pi}{2}$ we can obtain
\begin{equation}
    H_{U_A}\simeq \int \mathrm{d}x\left(\frac{U_A}{2\pi^2}\left(\left(\partial_x\phi_c\right)^2-\left(\partial_x\phi_s\right)^2\right)+\frac{U_A}{2\pi^2\alpha^2}\left(\cos\left(2\sqrt{2}\phi_s\right)+\cos\left(2\sqrt{2}\phi_c\right)\right)\right).
\end{equation}
We note that for $U$, i.e., on-site Hubbard interaction, the bosonized form at half filling is well known as 
\begin{equation}
    H_{U}\simeq \int \mathrm{d}x\left(\frac{U}{2\pi^2}\left(\left(\partial_x\phi_c\right)^2-\left(\partial_x\phi_s\right)^2\right)+\frac{U}{2\pi^2\alpha^2}\left(\cos\left(2\sqrt{2}\phi_s\right)-\cos\left(2\sqrt{2}\phi_c\right)\right)\right).
\end{equation}
For repulsive interactions, both terms renormalize the charge Luttinger parameter toward $K_c<1$, making the umklapp perturbation relevant. The crucial distinction lies in the opposite signs of the charge cosine. For the projected $A$-orbital interaction, the minima satisfy
\begin{equation}
    2\sqrt{2}\phi_c=(2n+1)\pi\Rightarrow\phi_c=\frac{(2n+1)\pi}{2\sqrt{2}},
\end{equation}
where $n\in\mathbb{Z}$. By contrast, the conventional on-site Hubbard interaction pins the charge field at
\begin{equation}
       2\sqrt{2}\phi_c=2n\pi\Rightarrow\phi_c=\frac{n\pi}{\sqrt{2}}.
\end{equation}
Thus, although both interactions gap the charge sector through a commensurate umklapp process, they pin $\phi_c$ at inequivalent sets of minima. The $U_A$ interaction therefore realizes a distinct Mott instability from that of the conventional 1D Hubbard model.

The density operators can read from Table.~\ref{tab:bosonization_dictionary}, and calculated at half filling as:
\begin{equation}
\begin{split}
n(x)\simeq&\sum_\sigma\left(-\frac{1}{\pi}\partial_x\phi_\sigma(x)
+
e^{2\mathrm{i}k_Fx}\psi^\dagger_{L;\sigma}\psi_{R;\sigma}
+
e^{-2\mathrm{i}k_Fx}\psi^\dagger_{R;\sigma}\psi_{L;\sigma}\right)\\
=&-\frac{\sqrt{2}}{\pi}\partial_x\phi_c(x)+\frac{2}{\pi\alpha}\sin\left(\sqrt{2}\phi_c(x)-2k_Fx\right)\cos\left(\sqrt{2}\phi_s(x)\right)\\
=&-\frac{\sqrt{2}}{\pi}\partial_x\phi_c(x)+\frac{2}{\pi\alpha}\sin\left(\sqrt{2}\phi_c(x)-\pi x\right)\cos\left(\sqrt{2}\phi_s(x)\right),
\end{split}
\end{equation}
For $\phi_c=\frac{(2n+1)\pi}{2\sqrt{2}}$, this factor has maximal magnitude and alternates from site to site. Open boundaries can therefore induce a pronounced $2k_F$ Friedel oscillation. By contrast, for the conventional Mott insulator, where $\phi_c=\frac{n\pi}{\sqrt{2}}$, the corresponding component vanishes at integer lattice positions. This difference provides a natural bosonization explanation for the Friedel oscillations observed in the DMRG calculations presented in the main text.
\begin{figure}[t]
\includegraphics[width=\columnwidth]{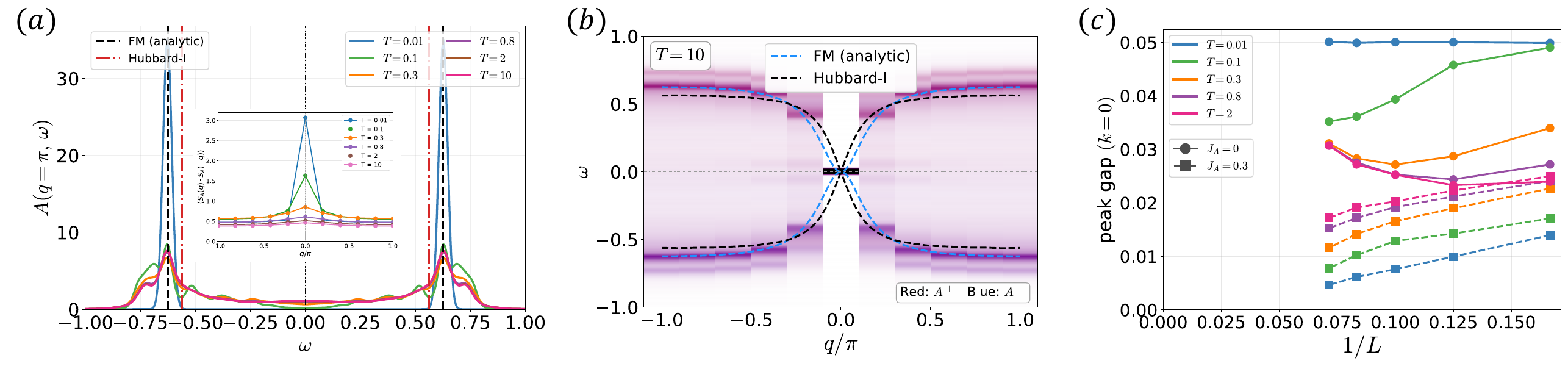}
\caption{\label{fig:ED_melt}
(a) We calculate the effective model at $k_*=0.5$. (a) Finite temperature single particle spectral function $A(k=\pi,\omega)$ for several temperatures. The dominant Hubbard-band peaks are consistent with the ferromagnetic analytic prediction, rather than the Hubbard-I approximation. The inset indicates that the ferromagnetic order is suppressed at high temperature. (b) Momentum-resolved spectral function $A(k,\omega)$ at $T = 10$. The dashed lines denote the ferromagnetic analytic result and the Hubbard-I prediction. (c) Finite size scaling of the spectral function peak gap at different temperatures. The gap vanishes finite upon extrapolation to the thermodynamic limit, indicating the stability in the presence of small $U = 0.05$.
}
\end{figure}
\section{Thermal melting of the ferromagnet}
In this section, we study the thermal melting of the model defined in the main text, which can be written as:
\begin{equation}
H_\mathrm{eff} =\sum_i \left(\frac{U_A}{2}\left(\tilde{n}_{A;i+\frac{1}{2}}-1\right)^2+\frac{U}{2}\left(n_{i}-1\right)^2 \right)+\sum_i \left(\frac{J_A}{2}\{\tilde{\mathbf{S}}_{A;i-\frac{1}{2}},\tilde{\mathbf S}_{A;i+\frac{1}{2}} \}+J\mathbf{S}_{i}\cdot \mathbf{S}_{i+1}\right).
    \label{eq:app_H}
\end{equation}
We focus on the half filling case with parameters $U_A=1,J=0$.  At $J_A=0$, the flat-band Stoner instability favors a fully spin-polarized ferromagnetic ground state. The FM phase can  be thermally suppressed and we may expect a thermal Mott semimetal as discussed in TBG~\cite{PhysRevX.15.021087,zhao2025ancilla,vituri2026controlledloopexpansiontopological,wei2026lifetimespectralfunctiontopological,nosov2026controlledexpansioncorrelatedelectrons}.
However, such a thermal Mott semimetal does not seem to exist in our model.  As shown in Fig.~\ref{fig:ED_melt}(b), once the ferromagnetic polarization is thermally reduced, the dominant peaks of the finite-temperature spectral function $A(k=\pi,\omega)$ nevertheless remain well described by the ferromagnetic analytic prediction, rather than by the Hubbard-I approximation. The same conclusion holds for the momentum-resolved spectrum $A(k,\omega)$ over the full Brillouin zone, as shown in Fig.~\ref{fig:ED_melt}(c). Thus, even after the ferromagnetic order is thermally melted, the single-particle spectrum retains the imprint of the flat-band ferromagnetic correlations, and the Hubbard-I approximation fails to capture the Hubbard band dispersion.

We note that the model with only the $U_A$ and $J_A$ term is fine tuned in the sense that the $k=0$ mode simply decouples. We thus add a small on-site Hubbard interaction $U$, which opens a gap at $k=0$ in the FM phase.  At $J_A=0$, we find that the gap survives in finite temperature regime. Therefore, we conclude that a stable gapless charge mode does not emerge by simply thermally melting the FM order in our model.  

In below we provide the definition of spectral function and some analytical calculations of it.
\begin{figure}[t]
\includegraphics[width=\columnwidth]{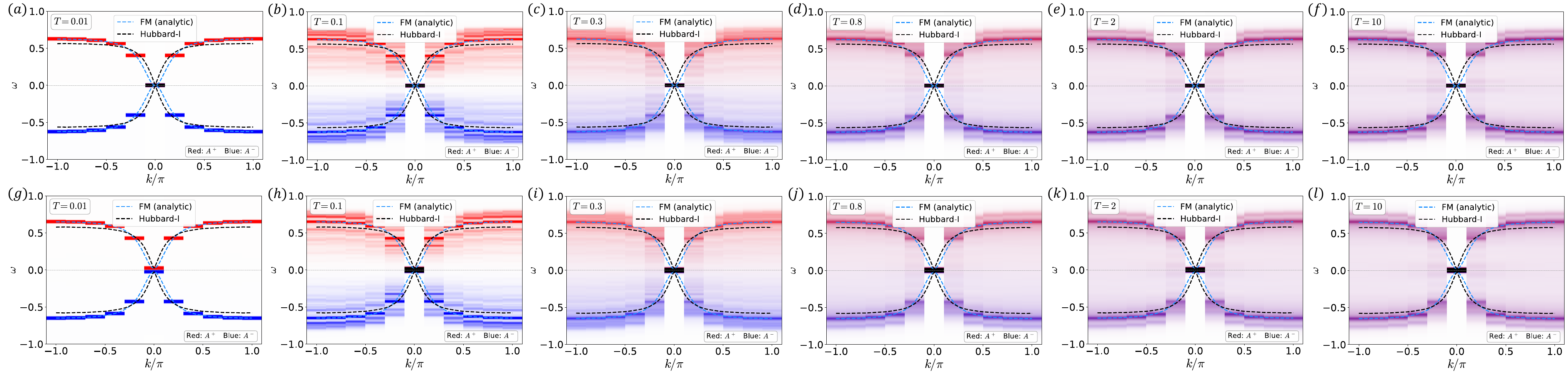}
\caption{\label{fig:spectralDifferentT}
Temperature evolution of $A(k,\omega)$.
In (a)-(f), we show $A(k,\omega)$ for $J_A=0$ at temperatures $T=0.01,0.1,0.3,0.8,2,$ and $10$, respectively. In (g)-(l), we show the corresponding spectra for finite antiferromagnetic exchange $J_A=0.3$ at the same temperatures. The blue dashed lines denote the analytic prediction of the fully polarized ferromagnet, while the white dashed lines denote the Hubbard-I approximation. For both values of $J_A$, the dominant Hubbard-band features remain closer to the ferromagnetic analytic dispersion than to the Hubbard-I prediction over a wide temperature range.}
\end{figure}
\subsection{Definiton of spectral function}
Here we provide the details of calculating the spectral function $A(k,\omega)$, which is defined as:
\begin{equation}
\begin{split}
    A(k,\omega)=&\sum_\sigma\Bigg(\frac{1}{Z}\sum_{\bra{m}\in N_e+1}\sum_{\ket{n}\in N_e} e^{-\frac{E_n}{T}}\langle m \lvert c^\dagger_{k;\sigma}\lvert n\rangle\delta(\omega-(E_m-E_n))\\
    &+\frac{1}{Z}\sum_{\bra{m}\in N_e-1}\sum_{\ket{n}\in N_e} e^{-\frac{E_n}{T}}\langle m \lvert c_{-k;\bar{\sigma}}\lvert n\rangle\delta(\omega-(E_n-E_m))\Bigg),
\end{split}
\end{equation}
where $Z$ is the partition function at particle number $N_e$: $Z=\sum_{\ket{n}\in N_e}e^{-\frac{E_n}{T}}$. The spectral function for  is shown in Fig.~\ref{fig:spectralDifferentT}. In the calculation we use gaussian function $\frac{1}{\sqrt{2\pi}\sigma}e^{-\frac{x^2}{2\sigma^2}}$ with $\sigma=0.01$ to approximate $\delta(x)$.

\subsection{Analytical calculation of spectral function}
Here we provide the analytical derivation of the single-particle spectral function discussed in the main text. At $t=0$, the projected effective Hamiltonian can be written as
\begin{equation}
\begin{split}
    H_\mathrm{eff}=&\sum_i \left(\frac{U_A}{2}\left(\tilde n_{A;i+\frac{1}{2}}-1\right)^2+\frac{U}{2}\left(n_{i}-1\right)^2\right)+\sum_i\left(\frac{J_A}{2}\{\tilde{\mathbf{S}}_{A;i-\frac{1}{2}},\tilde{\mathbf{S}}_{A;i+\frac{1}{2}}\}+J \mathbf{S}_{i}\cdot \mathbf{S}_{i+1}\right)\\
    =&\frac{U_A}{L}\sum_{k,k^\prime,q}g_{k+q}g_k g_{k^\prime-q}g_{k^\prime}c^\dagger_{k+q;\uparrow}c_{k;\uparrow}c^\dagger_{k^\prime-q;\downarrow}c_{k^\prime;\downarrow}-\frac{U_A}{2}\sum_{k;\sigma}g_k^2c^\dagger_{k;\sigma}c_{k;\sigma}\\
    &+\frac{U}{L}\sum_{k,k^\prime,q}c^\dagger_{k+q;\uparrow}c_{k;\uparrow}c^\dagger_{k^\prime-q;\downarrow}c_{k^\prime;\downarrow}-\frac{U}{2}\sum_{k;\sigma}c^\dagger_{k;\sigma}c_{k;\sigma}\\
    &+\frac{J_A}{L}\sum_{k,k^\prime,q}\cos q g_{k+q}g_k g_{k^\prime-q}g_{k^\prime} \sum_{\alpha,\beta,\alpha^\prime,\beta^\prime} c^\dagger_{k+q;\alpha} c_{k;\beta} c^\dagger_{k^\prime-q;\beta^\prime} c_{k^\prime;\alpha^\prime}\boldsymbol{\sigma}_{\alpha\beta}\cdot\boldsymbol{\sigma}_{\beta^\prime\alpha^\prime}\\
    &+\frac{J}{L}\sum_{k,k^\prime,q} \cos q\sum_{\alpha,\beta,\alpha^\prime,\beta^\prime}c^\dagger_{k+q;\alpha} c_{k;\beta} c^\dagger_{k^\prime-q;\beta^\prime} c_{k^\prime;\alpha^\prime}\boldsymbol{\sigma}_{\alpha\beta}\cdot\boldsymbol{\sigma}_{\beta^\prime\alpha^\prime}+\mathrm{const.},
\end{split}
\end{equation}
where $g_k=\frac{2\sin\frac{k}{2}}{\sqrt{(k^2_*+4\sin^2\frac{k}{2})\bar{z}}}$, $\bar{z}=\frac{1}{L}\sum_k\frac{4\sin^2\frac{k}{2}}{k_*^2+4\sin^2\frac{k}{2}}$ as defined in the main text. Below we first evaluate the spectral function in a fully polarized ferromagnetic state, and then derive the Hubbard-I approximation result.
\subsubsection{Fully polarized FM state}
We consider the case $J_A=J=0$. Without loss of generality, we choose the fully polarized state as $\lvert\mathrm{FM}\rangle=\prod_i c^\dagger_{i;\downarrow}\lvert0\rangle$. Since all spin-$\downarrow$ orbitals are occupied, we have
\begin{equation}
    \langle \mathrm{FM}\lvert c^\dagger_{k^\prime-q;\downarrow}c_{k^\prime;\downarrow}\lvert\mathrm{FM}\rangle=\delta_{q,0}.
\end{equation}
Therefore, adding a spin-$\uparrow$ electron, or equivalently removing a spin-$\downarrow$ electron by particle-hole symmetry at $t=0$, costs the energy $\frac{U_A}{2}g_k^2+\frac{U}{2}$. Based on the particle-hole symmetry at $t=0$. The spectral function at $T=0$ can be written as
\begin{equation}
    A(k,\omega)_{\mathrm{FM}}=\delta\left(\omega-\left(\frac{U_A}{2}g_k^2+\frac{U}{2}\right)\right)+\delta\left(\omega+\left(\frac{U_A}{2}g_k^2+\frac{U}{2}\right)\right).
\end{equation}
\subsubsection{Hubbard-I approximation}
We now derive the Hubbard-I approximation for the spectral function. The equation of motion for the single-particle Green's function is
\begin{equation}\label{eq:EOMG}
\begin{split}
    \omega \langle\!\langle c_{k;\sigma};c^\dagger_{k;\sigma}\rangle\!\rangle_\omega=&\langle\{c_{k;\sigma},c^\dagger_{k;\sigma}\}\rangle+\langle\!\langle [c_{k;\sigma},H_\mathrm{eff}];c^\dagger_{k;\sigma}\rangle\!\rangle_\omega\\
    =&1+\langle\!\langle [c_{k;\sigma},H_\mathrm{eff}];c^\dagger_{k;\sigma}\rangle\!\rangle_\omega,
\end{split}
\end{equation}
For $J_A=J_B=0$, the commutator is
\begin{equation}\label{eq:eta}
\begin{split}
    [c_{k;\sigma},H_\mathrm{eff}]=&\sum_{k^\prime,q}\left(\frac{U_A}{L}g_k g_{k-q} g_{k^\prime-q} g_{k^\prime}+\frac{U}{L}\right)c_{k-q;\sigma}c^\dagger_{k^\prime-q;\bar{\sigma}}c_{k^\prime;\bar{\sigma}}-\left(\frac{U_A}{2}g_k^2+\frac{U}{2}\right)c_{k;\sigma},
\end{split}
\end{equation}
where $\bar\sigma$ denotes the spin opposite to $\sigma$.
We define
\begin{equation}\label{eq:defOFGF}
\begin{split}
    G_\sigma(k,\omega)=&\langle\!\langle c_{k;\sigma};c^\dagger_{k;\sigma}\rangle\!\rangle_\omega,\\
    \eta_{k;\sigma}=&[c_{k;\sigma},H_\mathrm{eff}],\\
    F_\sigma(k,\omega)=&\langle\!\langle\eta_{k;\sigma};c^\dagger_{k;\sigma}\rangle\!\rangle_\omega.
\end{split}
\end{equation}
The equation of motion for $F_\sigma$ is
\begin{equation}\label{eq:Fequation}
    \omega F_\sigma(k,\omega)=\langle\{\eta_{k;\sigma},c^\dagger_{k;\sigma}\}\rangle+\langle\!\langle[\eta_{k;\sigma},H_\mathrm{eff}];c^\dagger_{k;\sigma}\rangle\!\rangle_\omega.
\end{equation}
At half filling, the anti-commutator is calculated as:
\begin{equation}\label{eq:ACetac}
    \langle\{\eta_{k;\sigma},c^\dagger_{k;\sigma}\}\rangle=\frac{1}{L}\sum_{k^\prime}\left(U_Ag_k^2g_{k^\prime}^2+U\right)\langle c^\dagger_{k^\prime;\bar{\sigma}}c_{k^\prime;\bar{\sigma}}\rangle-\left(\frac{U_A}{2}g^2_k+\frac{U}{2}\right)=0.
\end{equation}
The Hubbard-I approximation closes the hierarchy by projecting the higher-order operator $[\eta_{k;\sigma},H_{\rm eff}]$ back to the single-particle operator $c_{k;\sigma}$:
\begin{equation}
[\eta_{k;\sigma},H_\mathrm{eff}]\approx M(k)c_{k;\sigma}.
\end{equation}
The coefficient $M(k)$ is fixed by the anticommutator,
\begin{equation}\label{eq:Mexpression}
    M(k)=\langle \{[\eta_{k;\sigma},H_\mathrm{eff}],c^\dagger_{k;\sigma}\}\rangle=\langle\{\eta_{k;\sigma},\eta^\dagger_{k;\sigma}\}\rangle.
\end{equation}
Writing $\eta_{k;\sigma}=\sum_q \rho_{k;\bar{\sigma}}(q)c_{k-q;\sigma}$, where $\rho_{k;\bar{\sigma}}(q)$ can be defined through Eq.~\ref{eq:eta}, which is
\begin{equation}
    \rho_{k;\bar{\sigma}}(q)=\sum_{k^\prime}\left(\frac{U_A}{L}g_k g_{k-q} g_{k^\prime-q} g_{k^\prime}+\frac{U}{L}\right)c^\dagger_{k^\prime-q;\bar{\sigma}}c_{k^\prime;\bar{\sigma}}-\left(\frac{U_A}{2}g_k^2+\frac{U}{2}\right)\delta_{q,0}.
\end{equation}
Then we obtain
\begin{equation}
\begin{split}
    \langle\{\eta_{k;\sigma},\eta^\dagger_{k;\sigma}\}\rangle=&\sum_{q,q^\prime}\langle\rho_{k;\bar{\sigma}}(q) c_{k-q;\sigma}c^\dagger_{k-q^\prime;\sigma}\rho^\dagger_{k;\bar{\sigma}}(q^\prime)+c^\dagger_{k-q^\prime;\sigma}\rho^\dagger_{k;\bar{\sigma}}(q^\prime)\rho_{k;\bar{\sigma}}(q) c_{k-q;\sigma}\rangle\\
    \approx&\sum_{q,q^\prime}\left(\langle c_{k-q;\sigma}c^\dagger_{k-q^\prime;\sigma}\rangle\langle\rho_{k;\bar{\sigma}}(q)\rho^\dagger_{k;\bar{\sigma}}(q^\prime)\rangle+\langle c^\dagger_{k-q^\prime;\sigma}c_{k-q;\sigma}\rangle\langle\rho^\dagger_{k;\bar{\sigma}}(q^\prime)\rho_{k;\bar{\sigma}}(q)\rangle\right)\\
    =&\frac{1}{2}\sum_q\langle\{\rho_{k;\bar{\sigma}}(q),\rho^\dagger_{k;\bar{\sigma}}(q)\}\rangle\\
    =&\frac{1}{2}\sum_q\sum_{k^\prime,k^{\prime\prime}}\left(\frac{U_A}{L}g_k g_{k-q} g_{k^\prime-q} g_{k^\prime}+\frac{U}{L}\right)\left(\frac{U_A}{L}g_k g_{k-q} g_{k^{\prime\prime}-q} g_{k^{\prime\prime}}+\frac{U}{L}\right)\\
    &\left(\langle c^\dagger_{k^\prime-q;\bar{\sigma}}c_{k^\prime;\bar{\sigma}}c^\dagger_{k^{\prime\prime};\bar{\sigma}}c_{k^{\prime\prime}-q;\bar{\sigma}}\rangle+\langle c^\dagger_{k^{\prime\prime};\bar{\sigma}}c_{k^{\prime\prime}-q;\bar{\sigma}}c^\dagger_{k^\prime-q;\bar{\sigma}}c_{k^\prime;\bar{\sigma}}\rangle\right)\\
    &-\sum_q \sum_{k^\prime}\left(\frac{U_A}{L}g_k g_{k-q} g_{k^\prime-q} g_{k^\prime}+\frac{U}{L}\right)\left(\frac{U_A}{2}g_k^2+\frac{U}{2}\right)\delta_{q,0}\langle c^\dagger_{k^\prime-q;\bar{\sigma}}c_{k^\prime;\bar{\sigma}}\rangle\\
    &+\frac{1}{2}\sum_q \left(\frac{U_A}{2}g_k^2+\frac{U}{2}\right)^2\delta_{q,0}.
\end{split}
\end{equation}
By using the Hubbard-I approximation, we have $\langle c^\dagger_{k^\prime-q;\bar{\sigma}}c_{k^\prime;\bar{\sigma}}\rangle_{\mathrm{HI}}=\frac{1}{2}\delta_{q,0}$ and $\langle c^\dagger_{k^{\prime\prime};\bar{\sigma}}c_{k^{\prime\prime}-q;\bar{\sigma}}c^\dagger_{k^\prime-q;\bar{\sigma}}c_{k^\prime;\bar{\sigma}}\rangle_\mathrm{HI}=\langle c^\dagger_{k^\prime-q;\bar{\sigma}}c_{k^\prime;\bar{\sigma}}c^\dagger_{k^{\prime\prime};\bar{\sigma}}c_{k^{\prime\prime}-q;\bar{\sigma}}\rangle_\mathrm{HI}=\frac{1}{4}(\delta_{q,0}+\delta_{k^\prime,k^{\prime\prime}})$ at half filling. Then we can get
\begin{equation}\label{eq:etaFinal}
\begin{split}
\langle\{\eta_{k;\sigma},\eta^\dagger_{k;\sigma}\}\rangle=&\frac{1}{4}\sum_{k^\prime,k^{\prime\prime}}\left(\frac{U_A}{L}g_k^2  g_{k^\prime}^2+\frac{U}{L}\right)\left(\frac{U_A}{L}g_k^2 g^2_{k^{\prime\prime}}+\frac{U}{L}\right)+\frac{1}{4}\sum_{k^\prime,q}\left(\frac{U_A}{L}g_k g_{k-q} g_{k^\prime-q} g_{k^\prime}+\frac{U}{L}\right)^2\\
&-\frac{1}{2}\sum_{k^\prime}\left(\frac{U_A}{L}g_k^2  g_{k^\prime}^2+\frac{U}{L}\right)\left(\frac{U_A}{2}+\frac{U}{2}\right)+\frac{1}{2}\left(\frac{U_A}{2}+\frac{U}{2}\right)^2\\
=&\frac{1}{4}\sum_{k^\prime,q}\left(\frac{U_A}{L}g_k g_{k-q} g_{k^\prime-q} g_{k^\prime}+\frac{U}{L}\right)^2.
\end{split}
\end{equation}
Combining Eq.~\ref{eq:Fequation},\ref{eq:ACetac},\ref{eq:Mexpression} and \ref{eq:etaFinal}
 together, we obtain:
 \begin{equation}
     \omega F_{\sigma}(k,\omega)=\frac{1}{4}\sum_{k^\prime,q}\left(\frac{U_A}{L}g_k g_{k-q} g_{k^\prime-q} g_{k^\prime}+\frac{U}{L}\right)^2\langle\!\langle c_{k;\sigma};c^\dagger_{k;\sigma}\rangle\!\rangle_\omega.
 \end{equation}
 Together with Eq.~\ref{eq:EOMG} and \ref{eq:defOFGF}, this gives:
 \begin{equation}
     \begin{split}
         \omega G_{\sigma}(k,\omega)=&1+F_\sigma(k,\omega),\\
         \omega F_\sigma(k,\omega)=&\frac{1}{4}\sum_{k^\prime,q}\left(\frac{U_A}{L}g_k g_{k-q} g_{k^\prime-q} g_{k^\prime}+\frac{U}{L}\right)^2G_\sigma(k,\omega).
     \end{split}
 \end{equation}
Therefore,
 \begin{equation}
     \begin{split}
         G_\sigma(k,\omega)=&\frac{1}{\omega-\Sigma(k,\omega)},\\
         \Sigma(k,\omega)=&\frac{1}{4\omega}\sum_{k^\prime,q}\left(\frac{U_A}{L}g_k g_{k-q} g_{k^\prime-q} g_{k^\prime}+\frac{U}{L}\right)^2.
     \end{split}
 \end{equation}
 The spectral function can be written as:
 \begin{equation}
  \begin{split}
     A(k,\omega)_\mathrm{HI}=&2\delta\left(\omega-\frac{1}{2}\sqrt{\sum_{k^\prime,q}\left(\frac{U_A}{L}g_k g_{k-q} g_{k^\prime-q} g_{k^\prime}+\frac{U}{L}\right)^2}\right)\\
     &+2\delta\left(\omega+\frac{1}{2}\sqrt{\sum_{k^\prime,q}\left(\frac{U_A}{ L}g_k g_{k-q} g_{k^\prime-q} g_{k^\prime}+\frac{U}{L}\right)^2}\right).
 \end{split}
 \end{equation}


  
\begin{figure}[t]
\includegraphics[width=\columnwidth]{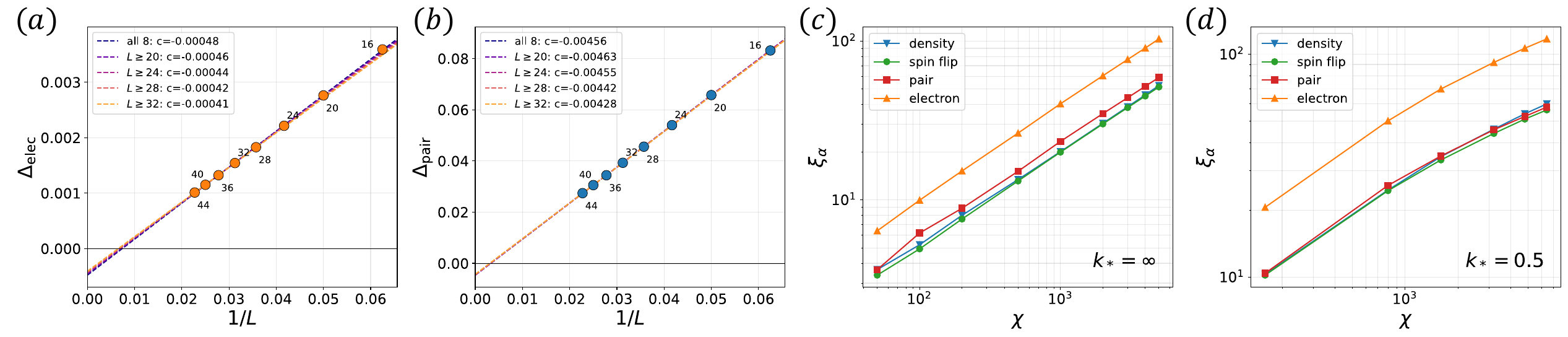}
\caption{\label{fig:finite_size}
Finite size and finite correlation length scaling in the Mott semimetal$^*$ regime. The parameters are $t=0,U_A=1,J_A=\frac{4}{3},U=\frac{1}{15},J=0$.
(a)(b) We calculate at $k_*=\infty$. Finite size scaling of the single-electron gap $\Delta_{\rm elec}$ and pair gap $\Delta_{\rm pair}$, respectively. Symbols denote finite DMRG results for different system sizes $L$, while dashed lines show linear fits in $\frac{1}{L}$ using different fitting windows. The extrapolated gaps are consistent with zero in the thermodynamic limit. The DMRG calculations are performed at bond dimension $\chi=5000$.
(c)($k_*=\infty,J_A=\frac{4}{3})$(d)($k_*=0.5,J_A=0.5$) iDMRG correlation lengths $\xi_\alpha$ as functions of bond dimension $\chi$ in the density, spin-flip, pair, and single-electron channels. The correlation lengths grow systematically with increasing $\chi$.
}
\end{figure}

 \section{Additional DMRG results}
In this section, we present additional DMRG results supporting the stability
of the Mott semimetal$^*$ phase. We perform finite DMRG and iDMRG calculations at
$t=0$, $U_A=1$, $J_A=\frac{4}{3}$,
$U=\frac{1}{15}$, $J=0$.
The results are shown in Fig.~\ref{fig:finite_size}.
Finite size scaling shows that both the single-electron gap and the pair gap
extrapolate to zero in the thermodynamic limit $L\rightarrow\infty$.
Consistently, in iDMRG the correlation lengths in the density, spin-flip,
pair, and single-electron channels all increase with bond dimension $\chi$,
supporting a gapless semimetallic state at finite $U$.


\begin{figure}[t]
\includegraphics[width=\columnwidth]{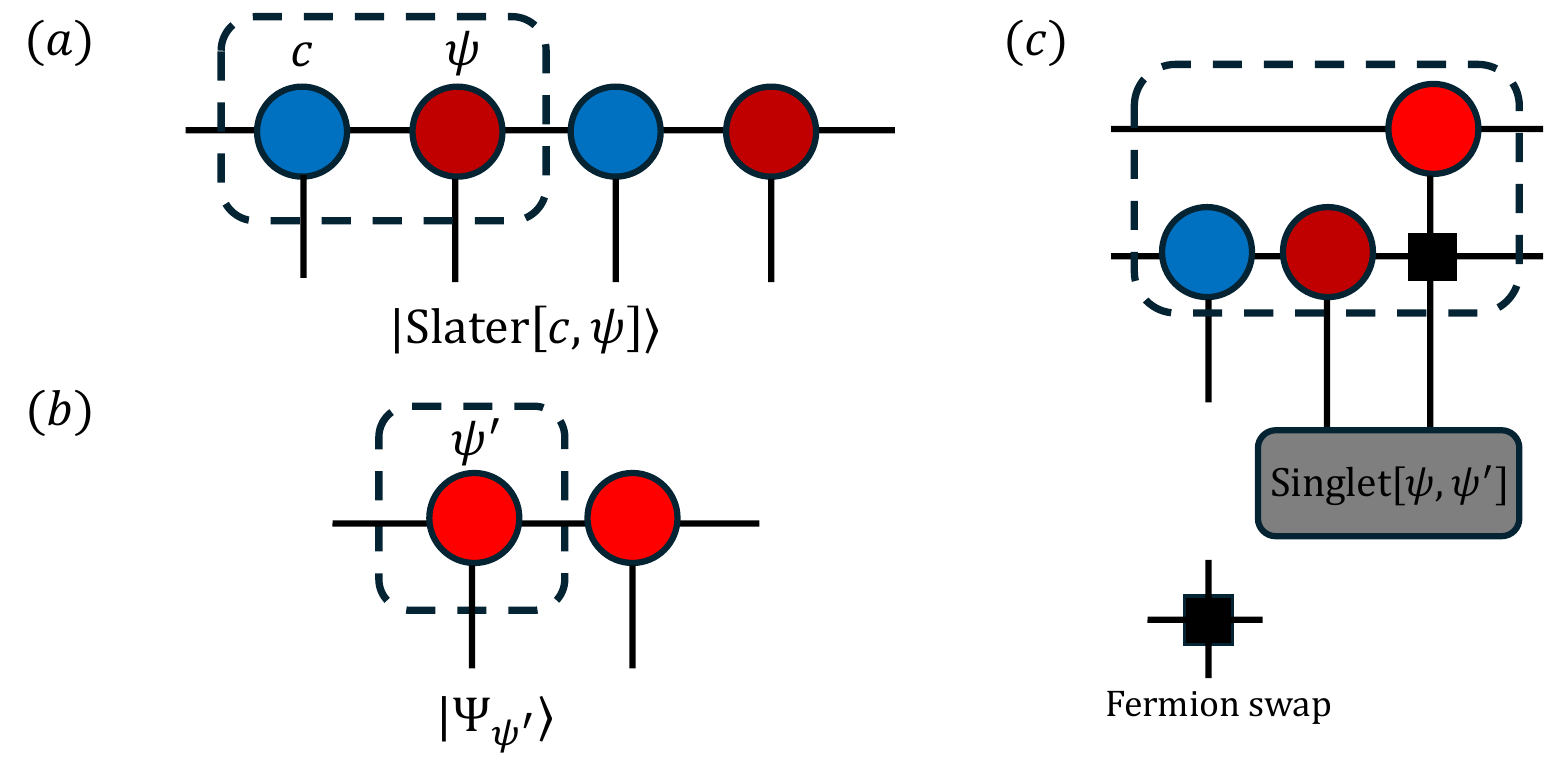}
\caption{\label{fig:Tensors}
MPS representation for the ancilla wavefunction. (a)(b) MPS representation for (a) $\lvert\mathrm{Slater}[c,\psi]\rangle$ and (b) $\lvert\Psi_{\psi^\prime}\rangle$. (c) Illustration of singlet projection in one unit cell. The fermion
swap is a diagonal tensor to correctly handle the fermionic anticommutation relations. $U_S$ denotes a local $SU(2)$ transformation that preserves the ancilla wavefunction.}
\end{figure}
\section{Construction of ancilla wavefunction}
In this section, we describe the construction of the ancilla wavefunction in the tensor-network language. As discussed in the main text, the final wavefunction is 
\begin{equation}
    \lvert\psi_\mathrm{ancilla}\rangle=P_S\lvert\mathrm{Slater}[c,\psi]\rangle\otimes \lvert\Psi_{\psi^\prime}\rangle,
\end{equation}
where $P_S$ projects the two ancilla fermions into a $SU(2)$ spin singlet at each site $i$. $\lvert\mathrm{Slater}[c,\psi]\rangle$ is given by a mean field Hamiltonian and $\lvert\Psi_{\psi^\prime}\rangle$ is the ground state of 1D Heisenberg model. For the $\mathbb{P}=\frac{1}{2}$ Mott insulator discussed in the main text, the physical electron $c$ is placed on site $i$, whereas the two ancilla fermions $\psi$ and $\psi^\prime$ are placed on the bond-centered site $i+\frac{1}{2}$. In the tensor-network representation, the singlet projection $P_S$ is implemented by contracting the physical legs associated with $\psi_{i+\frac{1}{2};\sigma}$ and $\psi^\prime_{i+\frac{1}{2};\sigma}$ with the local spin-singlet tensor at each bond-centered site. After this contraction, all ancilla physical legs are eliminated, and the resulting MPS contains only the physical legs corresponding to the electrons $c_{i;\sigma}$.
\begin{table*}[t]
\centering
\small
\renewcommand{\arraystretch}{1.35}
\setlength{\tabcolsep}{6pt}
\begin{tabular}{@{}p{0.22\textwidth}p{0.74\textwidth}@{}}
\toprule
Object & Bosonized expression \\
\midrule

Lattice fermions
&
$\displaystyle
c_{i;\sigma}
\simeq
\frac{R_\sigma+L_\sigma}{\sqrt{2}},
\quad
\psi_{i;\sigma}
\simeq
-\frac{\mathrm{i}}{\sqrt{2}}
\left(R_\sigma-L_\sigma\right)
$
\\

Chiral fermion
&
$\displaystyle
r_\sigma(x)
=
\frac{\kappa_\sigma}{\sqrt{2\pi\alpha}}\,
e^{-\mathrm{i}[r\phi_\sigma(x)-\theta_\sigma(x)]},
\quad
r=+1\;(R),\,-1\;(L),
\quad
\{\kappa_\sigma,\kappa_{\sigma'}\}
=
2\delta_{\sigma\sigma'}
$
\\

Charge and spin fields
&
$\displaystyle
\phi_{c,s}
=
\frac{\phi_\uparrow\pm\phi_\downarrow}{\sqrt{2}},
\quad
\theta_{c,s}
=
\frac{\theta_\uparrow\pm\theta_\downarrow}{\sqrt{2}}
$
\\

Same spin bilinears
&
$\displaystyle
:R^\dagger_\sigma R_\sigma:
=
-\frac{\partial_x\phi_\sigma-\partial_x\theta_\sigma}{2\pi},
\quad
:L^\dagger_\sigma L_\sigma:
=
-\frac{\partial_x\phi_\sigma+\partial_x\theta_\sigma}{2\pi}
$
\\

&
$\displaystyle
R^\dagger_\sigma L_\sigma
=
-\frac{\mathrm{i}}{2\pi\alpha}e^{2\mathrm{i}\phi_\sigma},
\quad
L^\dagger_\sigma R_\sigma
=
\frac{\mathrm{i}}{2\pi\alpha}e^{-2\mathrm{i}\phi_\sigma}
$
\\
Spin flip binears & $\displaystyle R^\dagger_\uparrow R_\downarrow=\frac{\mathrm{i}}{2\pi\alpha}e^{\mathrm{i}\sqrt{2}(\phi_s-\theta_s)},\quad R^\dagger_\downarrow R_\uparrow=-\frac{\mathrm{i}}{2\pi\alpha}e^{-\mathrm{i}\sqrt{2}(\phi_s-\theta_s)} $\\
 & $\displaystyle L^\dagger_\uparrow L_\downarrow=\frac{\mathrm{i}}{2\pi\alpha}e^{-\mathrm{i}\sqrt{2}(\phi_s+\theta_s)},\quad L^\dagger_\downarrow L_\uparrow=-\frac{\mathrm{i}}{2\pi\alpha}e^{\mathrm{i}\sqrt{2}(\phi_s+\theta_s)}\quad $\\
 & $\displaystyle R^\dagger_\uparrow L_\downarrow =\frac{1}{2\pi\alpha}e^{\mathrm{i}\sqrt{2}(\phi_c-\theta_s)},\quad L^\dagger_\downarrow R_\uparrow =\frac{1}{2\pi\alpha}e^{-\mathrm{i}\sqrt{2}(\phi_c-\theta_s)} $\\
 & $\displaystyle R^\dagger_\downarrow L_\uparrow =-\frac{1}{2\pi\alpha}e^{\mathrm{i}\sqrt{2}(\phi_c+\theta_s)},\quad L^\dagger_\uparrow R_\downarrow =-\frac{1}{2\pi\alpha}e^{-\mathrm{i}\sqrt{2}(\phi_c+\theta_s)} $\\
 Spin fields
&
$\displaystyle
\mathbf{J}_r
=
\frac{1}{2}
r^\dagger_\alpha
\boldsymbol{\sigma}_{\alpha\beta}
r_\beta,
\quad
\mathbf{N}
=
\frac{1}{2}
R^\dagger_\alpha
\boldsymbol{\sigma}_{\alpha\beta}
L_\beta
$
\\

Orbital densities
&
$\displaystyle
\begin{aligned}
n_{c;\sigma}
&=
\frac{1}{2}
\bigl(
R^\dagger_\sigma R_\sigma
+L^\dagger_\sigma L_\sigma
+R^\dagger_\sigma L_\sigma
+L^\dagger_\sigma R_\sigma
\bigr), \quad
n_{\psi;\sigma}
=
\frac{1}{2}
\bigl(
R^\dagger_\sigma R_\sigma
+L^\dagger_\sigma L_\sigma
-R^\dagger_\sigma L_\sigma
-L^\dagger_\sigma R_\sigma
\bigr).
\end{aligned}
$
\\

Orbital spin densities
&
$\displaystyle
\begin{aligned}
\mathbf{S}_c
&=
\frac{1}{2}
\left(\mathbf{J}_R+\mathbf{J}_L\right)
+
\frac{1}{2}
\left(\mathbf{N}+\mathbf{N}^\dagger\right),
\quad
\mathbf{S}_\psi
=
\frac{1}{2}
\left(\mathbf{J}_R+\mathbf{J}_L\right)
-
\frac{1}{2}
\left(\mathbf{N}+\mathbf{N}^\dagger\right).
\end{aligned}
$
\\

\bottomrule
\end{tabular}
\caption{Bosonization dictionary for $c$ and $\psi$ used in Appendix.~\ref{app:bosonization_ancilla}.}
\label{tab:bosonization_dictionary_ancilla}
\end{table*}

\begin{table*}[t]

\centering
\small
\renewcommand{\arraystretch}{1.35}
\setlength{\tabcolsep}{6pt}

\begin{tabular}{@{}p{0.22\textwidth}p{0.74\textwidth}@{}}
\toprule
Object & Bosonized expression \\
\midrule

Spin decomposition
&
$\displaystyle
\mathbf{S}_{\psi^\prime}
=\mathbf{J}^\prime_R+\mathbf{J}^\prime_L+
C(-1)^x\mathbf{N}^\prime,
$
\\

Spin operators
&
$\displaystyle
J_R^{\prime z}
=
-\frac{1}{2\sqrt{2}\pi}\left(\partial_x\phi-\partial_x\theta\right),\quad J^{\prime z}_L =
-\frac{1}{2\sqrt{2}\pi}\left(\partial_x\phi+\partial_x\theta\right)$ \\

& $\displaystyle J^{\prime \pm}_R=\frac{1}{2\pi\alpha} e^{\pm\mathrm{i}\sqrt{2}(\phi-\theta)},\quad J^{\prime \pm}_L=\frac{1}{2\pi\alpha} e^{\mp\mathrm{i}\sqrt{2}(\phi+\theta)}
$
\\
&
$\displaystyle
N^{\prime z}=\cos(\sqrt{2}\phi),
\quad
N^{\prime \pm}=e^{\mp\mathrm{i}\sqrt{2}\theta}
$
\\

\bottomrule
\end{tabular}
\caption{Bosonization dictionary for $\mathbf{S}_{\psi^\prime}$ in Appendix.~\ref{app:bosonization_ancilla}.}
\label{tab:bosonization_dictionary_heisenberg}
\end{table*}

\section{Bosonization analysis of ancilla theory}\label{app:bosonization_ancilla}
The ancilla theory for this system can be described as:
\begin{equation}
\mathcal{L}=\mathcal{L_\mathrm{charge}}+\mathcal{L_\mathrm{spin}},
\end{equation}
where $\mathcal L_{\mathrm{spin}}$ represents the SU(2)$_1$ CFT for the neutral spin mode from $\psi^\prime$.  $\mathcal{L}_{\mathrm{charge}}$ represents a C1S1 phase for the itinerant sector with a spinful Dirac fermion from $c$ and $\psi$, captured by an effective low energy Hamiltonian as:
\begin{equation}
    \mathcal H_{\mathrm{charge}}=\sum_{k,\sigma} 
    \begin{pmatrix}
        c^\dagger_{k;\sigma} & \psi^\dagger_{k;\sigma}
    \end{pmatrix} (\upsilon k \tau^y + m \tau^z) \begin{pmatrix} c_{k;\sigma} \\ \psi_{k;\sigma}\end{pmatrix}.
\end{equation}
The $m$ term vanishes when particle-hole symmetry is preserved. 
To analyze the stability of the $\mathcal{L}_\mathrm{charge}$, we bosonize the operators $c$ and $\psi$ as:
\begin{equation}
    \begin{split}
       R_{\sigma}(x)\simeq&\frac{1}{\sqrt{2}}\left(c_{i;\sigma}+\mathrm{i}\psi_{i;\sigma}\right), \\
       L_\sigma(x) \simeq &\frac{1}{\sqrt{2}}\left(c_{i;\sigma}-\mathrm{i}\psi_{i;\sigma}\right).
    \end{split}
\end{equation}
Then we can bosonize other operators, the details are shown in Table.~\ref{tab:bosonization_dictionary_ancilla}.
\subsection{Symmetries and allowed perturbations}
The symmetry transformations of the fields are:
\begin{equation}
    \begin{split}
        \mathcal I:
\quad&
\phi_{c}(x)\rightarrow-\phi_{c}(-x)+\frac{\pi}{2},\quad
\theta_{c}(x)\rightarrow\theta_{c,s}(-x),\\
&\phi_s\rightarrow-\phi_{s}(-x),\quad
\theta_{s}(x)\rightarrow\theta_{s}(-x),\\
\mathcal C:
\quad
&\phi_{c,s}(x)\rightarrow-\phi_{c,s}(x),
\quad
\theta_{c,s}(x)\rightarrow\theta_{c,s}(x),\\
\mathcal T:
\quad
&\phi_c(x)\rightarrow\phi_c(x),
\quad
\theta_c(x)\rightarrow-\theta_c(x)-\frac{\pi}{2},
\\
&\phi_s(x)\rightarrow-\phi_s(x),
\quad
\theta_s(x)\rightarrow\theta_s(x)+\frac{\pi}{2}.
 \end{split}
\end{equation}
The shifts are defined modulo the compactification periods. Under above symmetry, the allowed $2$- and $4$-fermions terms in charge sectors $c,\psi$ are listed as:
\begin{equation}
\begin{split}
\mathcal{O}_{2,1}=&\left(\partial_x \phi_c\right)^2,\mathcal{O}_{2,2}=\left(\partial_x \theta_c\right)^2,\mathcal{O}_{2,3}=\left(\partial_x \phi_s\right)^2,\mathcal{O}_{2,4}=\left(\partial_x \theta_s\right)^2,\\
    \mathcal{O}_{4,1}=&\cos \left(2\sqrt{2}\phi_c\right), \mathcal{O}_{4,2}=\cos \left(2\sqrt{2}\phi_s\right),\\
    \mathcal{O}_{4,3}=&\left(\partial_x\phi_c\right)\sin\left(\sqrt{2}\phi_c\right)\cos \left(\sqrt{2}\phi_s\right),\mathcal{O}_{4,4}=\left(\partial_x\phi_s\right)\cos\left(\sqrt{2}\phi_c\right)\sin \left(\sqrt{2}\phi_s\right).
\end{split}
\end{equation}
We note that compared to the $k_F\ne0$ case, here terms like $R^\dagger R R^\dagger L$ are allowed.
The Gaussian terms can be absorbed in the Luttinger parameter $K_c$, here $K_s=1$ due to the spin $SU(2)$ symmetry. The Guassian action implies 
\begin{equation}
    \Delta[e^{\mathrm{i}\beta\phi_\nu}]=\frac{\beta^2 K_\nu}{4},\quad \Delta[e^{\mathrm{i}\beta\theta_\nu}]=\frac{\beta^2 }{4K_\nu},\quad
    \Delta[\partial_x\phi_\nu]=\Delta[\partial_x\theta_\nu]=1.
\end{equation}
Hence the RG equation at tree level can be obtained as:
\begin{equation}
    \begin{split}
       \frac{\mathrm{d} g_{4,1}}{\mathrm{d} l}=&(2-2K_c)g_{4,1},\\
       \frac{\mathrm{d} g_{4,2}}{\mathrm{d} l}=&0,\\
         \frac{\mathrm{d} g_{4,3}}{\mathrm{d} l}=&\frac{1-K_c}{2}g_{4,3},\\
         \frac{\mathrm{d} g_{4,4}}{\mathrm{d} l}=&\frac{1-K_c}{2}g_{4,4}.
        \end{split}
\end{equation}
The vanishing $\beta$ equation for $g_{4,2}$ indicates that we need one-loop correction calculation.
\subsection{Momentum-shell integration}
After integrating out $\theta_\nu$ and setting $y_\nu=v_\nu\tau$ where $v_\nu$ is charge/spin velocity. The Euclidean action is
\begin{equation}
    S=\sum_\nu \frac{1}{2\pi K_\nu}\int\mathrm{d}^2r\,(\nabla\phi_\nu)^2.
\end{equation}
Split $\phi_\nu=\phi_{\nu,<}+\phi_{\nu,>}$ with
\begin{equation}
 \phi_<:|q|<\Lambda e^{-\mathrm{d}\ell},\quad
 \phi_>:\Lambda e^{-\mathrm{d}\ell}<|q|<\Lambda,\quad
 \Lambda=\alpha^{-1}.
\end{equation}
The fast-mode fluctuations can be calculated as
\begin{equation}
 \langle\phi_{\nu,>}^2\rangle_>
 =\int_{\Lambda e^{-\mathrm{d}\ell}}^\Lambda\frac{q\,\mathrm{d} q}{(2\pi)^2}
 \int_0^{2\pi}\mathrm{d}\theta\,\frac{\pi K_\nu}{q^2}
 =\frac{K_\nu}{2}\mathrm{d}\ell.
\end{equation}
Consequently,
\begin{equation}
\langle e^{\mathrm{i}\beta\phi_\nu}\rangle_>
 =e^{\mathrm{i}\beta\phi_{\nu,<}}e^{-\beta^2K_\nu\mathrm{d}\ell/4}.
\end{equation}
Restoring the cutoff by $r=e^{\mathrm{d}\ell}r'$ contributes $e^{2\mathrm{d}\ell}$ from the measure and $e^{-\mathrm{d}\ell}$ per derivative. 
For terms $S_\mathrm{int}=\frac{g_{4,2}}{v_s}\int\mathrm{d}^2r\cos\left(2\sqrt{2}\phi_s\right)$, the cumulant expansion reads
\begin{equation}
 S_{\rm eff}=S_0+\langle S_{\rm int}\rangle_>
 -\frac12(\langle S_{\rm int}^2\rangle_>-\langle S_{\rm int}\rangle_>^2)+O(g_{4,2}^3).
\end{equation}
At two points $x_{1,2}=X\pm\frac{\rho}{2}$, the operator product expansion (OPE) gives that
\begin{align}
 &\cos\left(2\sqrt2\phi_s(X+\rho/2)\right)\cos\left(2\sqrt2\phi_s(X-\rho/2)\right)\\
 &\quad=\frac12\left(\frac{\alpha}{|\rho|}\right)^{4K_s}
 \left(1-4\rho_\mu\rho_\nu\partial_\mu\phi_s\partial_\nu\phi_s+\cdots\right)
 +\frac12\left(\frac{|\rho|}{\alpha}\right)^{4K_s}
 \cos(4\sqrt2\phi_s)+\cdots.
\end{align}
Integrating $\alpha<|\rho|<\alpha e^{\mathrm{d}\ell}$ yields the Kosterlitz-Thouless (KT) equations
\begin{equation}
 \frac{\mathrm{d}g_{4,2}}{\mathrm{d}\ell}=2(1-K_s)g_{4,2},\quad
 \frac{\mathrm{d} K_s}{\mathrm{d}\ell}=-\frac12K_s^2\left(\frac{v_s}{2\pi\alpha}\right)^2g_{4,2}^2.
\end{equation}
Therefore we obtain:
\begin{equation}
    \frac{\mathrm{d} g_{4,2}}{\mathrm{d} l}= -\frac{2\pi\alpha^2}{v_s}g^2_{4,2},
\end{equation}
So without $\mathcal{L}_\mathrm{spin}$, the Dirac fermion part is stable if $K_c>1,g_{4,2}>0$. For the instability, a possible microscopic origin would be an interaction term
\begin{equation}\label{eqn:Hintancilla}
    H_\mathrm{int}=\frac{U}{2}\left(n_c^2+n_\psi^2\right)+Vn_cn_\psi.
\end{equation}
Here $K_c<1$ happens when $U+V>0$ and $g_{4,2}<0$ happens when $U-V<0$.
\subsection{Coupling to the neutral spin mode}
If we consider the coupling to the neutral spin mode $\psi^\prime$, the most general coupling is
\begin{equation}\label{eqn:HJancilla}
    H_J=J_c\mathbf{S}_c\cdot \mathbf{S}_{\psi^\prime}+J_\psi \mathbf{S}_\psi\cdot \mathbf{S}_{\psi^\prime}.
\end{equation}
We bosonize the spin field of $\psi^\prime$, the details are given in Table.~\ref{tab:bosonization_dictionary_heisenberg}. Terms containing $(-1)^i N^\prime$ oscillate and average to zero in the low energy theory. The IR part is
\begin{equation}
\begin{split}
    \mathcal{H}^{\rm IR}_{J}=&\int \mathrm{d}x\left(g_{J,\parallel}\mathcal{O}_{J,\parallel}+g_{J,\perp}\mathcal{O}_{J,\perp}+g_{N}\mathcal{O}_{N}\right),\\
    \mathcal{O}_{J,\parallel}=&\mathbf{J}_R\cdot\mathbf{J}_R^\prime+\mathbf{J}_L\cdot\mathbf{J}_L^\prime,\\
    \mathcal{O}_{J,\perp}=&\mathbf{J}_R\cdot\mathbf{J}_L^\prime+\mathbf{J}_L\cdot\mathbf{J}_R^\prime,\\
    \mathcal{O}_{N}=&\left(\mathbf{N}+\mathbf{N}^\dagger\right)\cdot \left(\mathbf{J}^\prime_R+\mathbf{J}^\prime_L\right),
\end{split}
\end{equation}
where $g_{J,\parallel}=g_{J,\perp}=\frac{J_c+J_\psi}{2},g_N=\frac{J_c-J_\psi}{2}$. Here we omit the factor from lattice to continuum model. The bosonization form of $\mathcal{O}_{J,\parallel},\mathcal{O}_{J,\perp}$ and $\mathcal{O}_N$ is
\begin{equation}
\begin{split}
    \mathcal{O}_{J,\parallel}=&\frac{\left(\partial_x\phi_s\right)\left(\partial_x\phi\right)+\left(\partial_x\theta_s\right)\left(\partial_x\theta\right)}{4\pi^2}+\frac{\cos\left(\sqrt{2}\left(\phi_s-\phi\right)\right)\sin\left(\sqrt{2}\left(\theta_s-\theta\right)\right)}{2\pi^ 2\alpha^2},\\
    \mathcal{O}_{J,\perp}=&\frac{\left(\partial_x\phi_s\right)\left(\partial_x\phi\right)-\left(\partial_x\theta_s\right)\left(\partial_x\theta\right)}{4\pi^2}+\frac{\cos\left(\sqrt{2}\left(\phi_s+\phi\right)\right)\sin\left(\sqrt{2}\left(\theta_s-\theta\right)\right)}{2\pi^2\alpha^2},\\
    \mathcal{O}_N=&-\frac{\cos \left(\sqrt{2}\phi_c\right)\sin\left(\sqrt{2}\phi_s\right)\left(\partial_x\phi\right)}{\sqrt{2}\pi^2\alpha}+\frac{\sin\left(\sqrt{2}\phi_c\right)\cos\left(\sqrt{2}\phi\right)\sin\left(\sqrt{2}\left(\theta_s-\theta\right)\right)}{\pi^2\alpha^2}.
\end{split}
\end{equation}
The dimension of the above operators are:
\begin{equation}
    \Delta[\mathcal{O}_{J,\parallel}]=\Delta[\mathcal{O}_{J,\perp}]=2,\quad \Delta[\mathcal{O}_N]=\frac{K_c+3}{2}.
\end{equation}
Therefore the RG equation at tree level is
\begin{equation}
    \begin{split}
        \frac{\mathrm{d} g_{J,\parallel}}{\mathrm{d}l }=&0,\\
        \frac{\mathrm{d} g_{J,\perp}}{\mathrm{d}l }=&0,\\
\frac{\mathrm{d} g_N}{\mathrm{d}l }=&\frac{1-K_c}{2}g_N.
    \end{split}
\end{equation}
To calculate the one-loop result, we need to calculate $\delta S^{(2)}=-\frac{1}{2}\langle S^2_J\rangle_{>,c}$, where $S_J=\int\mathrm{d}^2r (g_{J,\parallel}\mathcal{O}_{J,\parallel}+g_{J,\perp}\mathcal{O}_{J,\perp}+g_{N}\mathcal{O}_{N})$. We define $\rho=(x,\tau)$, $z=x+\mathrm{i}v_s\tau$ and $z^\prime=x+\mathrm{i}v^\prime_s\tau$, where $v^\prime_s$ is the spin velocity of $\psi^\prime$. Using the $SU(2)_1$ current algebra
\begin{equation}\label{eqn:J_algebra}
    J^a_R(\rho)J^b_R(0)\sim \frac{\delta^{ab}}{8\pi^2z^2}+\frac{\mathrm{i}\epsilon^{abc}}{2\pi z}J^c_R(0),\quad J^a_L(\rho)J^b_L(0)\sim \frac{\delta^{ab}}{8\pi^2\bar{z}^2}+\frac{\mathrm{i}\epsilon^{abc}}{2\pi \bar{z}}J^c_L(0),
\end{equation}
we can calculate the correction from $\mathcal{O}_{J,\perp}^2,\mathcal{O}_{J,\perp}\mathcal{O}_N$ terms. For the terms including $\mathcal{O}_{J,\parallel}$, it does not contribute to the result since the same chirality integral vanishes. Here we provide the details of calculations for all the relevant terms.
\subsubsection{$\mathcal{O}_{J,\perp}^2$}
At two points $x_{1,2}=X\pm\frac{\rho}{2}$, correspondingly $z_{1,2}=Z\pm\frac{\zeta}{2}$ and $z^\prime_{1,2}=Z^\prime\pm\frac{\zeta^\prime}{2}$. Using Eq.~\ref{eqn:J_algebra}, we have
\begin{equation}
    \mathcal{O}_{J,\perp}(X+\frac{\rho}{2})\mathcal{O}_{J,\perp}(X-\frac{\rho}{2})\sim -\frac{\epsilon^{abc}\epsilon^{abd}}{4\pi^2\zeta\bar{\zeta}^\prime}J_R^c(x_2)J_L^{\prime d}(x_2) -\frac{\epsilon^{abc}\epsilon^{abd}}{4\pi^2\zeta^\prime\bar{\zeta}}J_R^{\prime c}(x_2)J_L^{d}(x_2)=-\frac{\mathbf{J}_R\cdot \mathbf{J}^\prime_L}{2\pi^2 \zeta\bar{\zeta}^\prime}-\frac{\mathbf{J}^\prime_R\cdot \mathbf{J}_L}{2\pi^2 \zeta^\prime \bar{\zeta}}.
\end{equation}
Integrating $\alpha<|\rho|<\alpha e^{\mathrm{d}\ell}$ yields
\begin{equation}\label{eqn:perpperp}
    -\frac{g^2_{J,\perp}}{2}\int \mathrm{d}^2 X\int_\mathrm{shell} \mathrm{d^2}\rho \mathcal{O}_{J,\perp}(X+\frac{\rho}{2})\mathcal{O}_{J,\perp}(X-\frac{\rho}{2})\sim \frac{g_{J,\perp}^2}{\pi(v_s+v_s^\prime)} \mathrm{d}\ell\int\mathrm{d}^2X\mathcal{O}_{J,\perp}(X).
\end{equation}
\subsubsection{$\mathcal{O}_{J,\perp}\mathcal{O}_N$}
Similarly, we can use Eq.~\ref{eqn:J_algebra} and calculate that
\begin{equation}
     \mathcal{O}_{J,\perp}(X+\frac{\rho}{2})\mathcal{O}_{N}(X-\frac{\rho}{2})\sim-\frac{1}{4\pi^2}\left(\frac{\left(\mathbf{N}+\mathbf{N}^\dagger\right)\cdot\mathbf{J}^\prime_L}{\zeta\bar{\zeta}^\prime}+\frac{\left(\mathbf{N}+\mathbf{N}^\dagger\right)\cdot\mathbf{J}^\prime_R}{\bar{\zeta}\zeta^\prime}\right).
\end{equation}
Integrating $\alpha<|\rho|<\alpha e^{\mathrm{d}\ell}$ yields
\begin{equation}\label{eqn:perpN}
    -g_{J,\perp}g_N\int \mathrm{d}^2 X\int_\mathrm{shell} \mathrm{d^2}\rho \mathcal{O}_{J,\perp}(X+\frac{\rho}{2})\mathcal{O}_{N}(X-\frac{\rho}{2})\sim \frac{g_{J,\perp}g_N}{\pi(v_s+v_s^\prime)} \mathrm{d}\ell\int\mathrm{d}^2X\mathcal{O}_{N}(X).
\end{equation}
\subsubsection{$\mathcal{O}_N^2$}
This channel contains the contraction of charge vertex. First we have
\begin{align}
N^a(\rho)N^{b\dagger}(0)
 \sim&\frac{\delta^{ab}}{8\pi^2\alpha\rho_s}
 \left(\frac{\alpha}{\rho_c}\right)^{K_c}+\frac{\mathrm{i}\epsilon^{abc}}{4\pi}
 \frac{\alpha^{K_c-1}}{\rho_c^{K_c}}
 \left(\frac{z_s}{\rho_s}J_R^c
       +\frac{\bar z_s}{\rho_s}J_L^c\right)+\cdots,\\
       N^{a\dagger}(\rho)N^{b}(0)
 \sim&\frac{\delta^{ab}}{8\pi^2\alpha\rho_s}
 \left(\frac{\alpha}{\rho_c}\right)^{K_c}+\frac{\mathrm{i}\epsilon^{abc}}{4\pi}
 \frac{\alpha^{K_c-1}}{\rho_c^{K_c}}
 \left(\frac{z_s}{\rho_s}J_R^c
       +\frac{\bar z_s}{\rho_s}J_L^c\right)+\cdots,
\label{eqn:N_algebra}
\end{align}
where $\rho_\nu=\sqrt{x^2+v_\nu^2\tau^2}$.
Using Eq.~\ref{eqn:J_algebra},\ref{eqn:N_algebra}, we can obtain
\begin{equation}
    \mathcal{O}_{N}(X+\frac{\rho}{2})\mathcal{O}_{N}(X-\frac{\rho}{2})\sim -\frac{\alpha^{K_c-1}}{2\pi^2\rho_c^{K_c}\rho_s}\left(\left(\frac{z}{z^\prime}\mathbf{J}_R\cdot \mathbf{J}^\prime_R+\frac{\bar{z}}{\bar{z}^\prime}\mathbf{J}_L\cdot \mathbf{J}^\prime_L\right)+\left(\frac{\bar{z}}{z^\prime}\mathbf{J}_L\cdot \mathbf{J}^\prime_R+\frac{z}{\bar{z}^\prime}\mathbf{J}_R\cdot \mathbf{J}^\prime_L\right)\right)+\cdots.
\end{equation}
$\cdots$ corresponds to other singular terms, here we only consider the correction to the operators $\mathcal{O}_{J,\parallel},\mathcal{O}_{J,\perp},\mathcal{O}_N$, therefore we neglect other terms. Integrating $\alpha<|\rho|<\alpha e^{\mathrm{d}\ell}$ yields
\begin{equation}\label{eqn:NN}
    -\frac{g_N^2}{2}\int \mathrm{d}^2 X\int_\mathrm{shell} \mathrm{d^2}\rho \mathcal{O}_{N}(X+\frac{\rho}{2})\mathcal{O}_{N}(X-\frac{\rho}{2})\sim g_N^2\left(\mathcal{C}_\parallel\mathcal{O}_{J,\parallel}+\mathcal{C}_\perp O_{J,\perp}\right),
\end{equation}
where 
\begin{align}
 \mathcal C_\parallel
 &=\frac{1}{4\pi^2v_s}\int_0^{2\pi}\mathrm{d}\theta\,
 \frac{\cos^2\theta+(v_s'/v_s)\sin^2\theta}
 {\left(\cos^2\theta+(v_s'/v_s)^2\sin^2\theta\right)
  \left(\cos^2\theta+(v_c/v_s)^2\sin^2\theta\right)^{K_c/2}},\\
 \mathcal C_\perp
 &=\frac{1}{4\pi^2v_s}\int_0^{2\pi}\mathrm{d}\theta\,
 \frac{\cos^2\theta-(v_s'/v_s)\sin^2\theta}
 {\left(\cos^2\theta+(v_s'/v_s)^2\sin^2\theta\right)
  \left(\cos^2\theta+(v_c/v_s)^2\sin^2\theta\right)^{K_c/2}}.
 \label{eq:Cparallel-Cperp}
\end{align}
One can verify that $\mathcal{C}_\parallel>0$ and $\mathcal{C}_\perp>0$ when $v_c>v_s$, $\mathcal{C}_\perp=0$ when $v_c=v_s$, $\mathcal{C}_\perp<0$ when $v_c<v_s$.
Combining Eq.~\ref{eqn:perpperp},\ref{eqn:perpN},\ref{eqn:NN}, we obtain the final RG equation at one-loop level:
\begin{equation}
    \begin{split}
        \frac{\mathrm{d} g_{J,\parallel}}{\mathrm{d} l} = &\mathcal{C}_\parallel g_N^2 ,\\
        \frac{\mathrm{d} g_{J,\perp}}{\mathrm{d} l} = & \frac{g^2_{J,\perp}}{\pi\left(v_s+v^\prime_s\right)}+\mathcal{C}_\perp g_N^2,\\
        \frac{\mathrm{d} g_{N}}{\mathrm{d} l}=&\left(\frac{1-K_c}{2}+\frac{g_{J,\perp}}{\pi(v_s+v^\prime_s)}\right)g_N.
    \end{split}
\end{equation}
The phase is stable when $K_c>1$ and $g_{J,\perp}<0$. Considering the microscopic interaction in Eq.~\ref{eqn:Hintancilla}\ref{eqn:HJancilla}, the final stable condition would be
\begin{equation}
    U+V<0,\quad U-V>0,\quad J_c+J_\psi<0.
\end{equation}
When an instability happens, there are several possibilities towards different phases:
\begin{itemize}
    \item For $U+V>0$, one has $K_c<1$, leading to a C0S2 or symmetry-breaking phase.

    \item For $U+V<0$ and $J_c+J_\psi>0$, the spin sectors of
    $c$, $\psi$, and $\psi'$ become unstable, driving the system into a
    C1S0 phase.

    \item For $U+V<0$, $J_c+J_\psi<0$ and $U-V<0$, only the spin
    sectors of $c$ and $\psi$ become unstable, driving the system into a
    C1S1 phase.
\end{itemize}
\begin{figure}[t]
\includegraphics[width=\columnwidth]{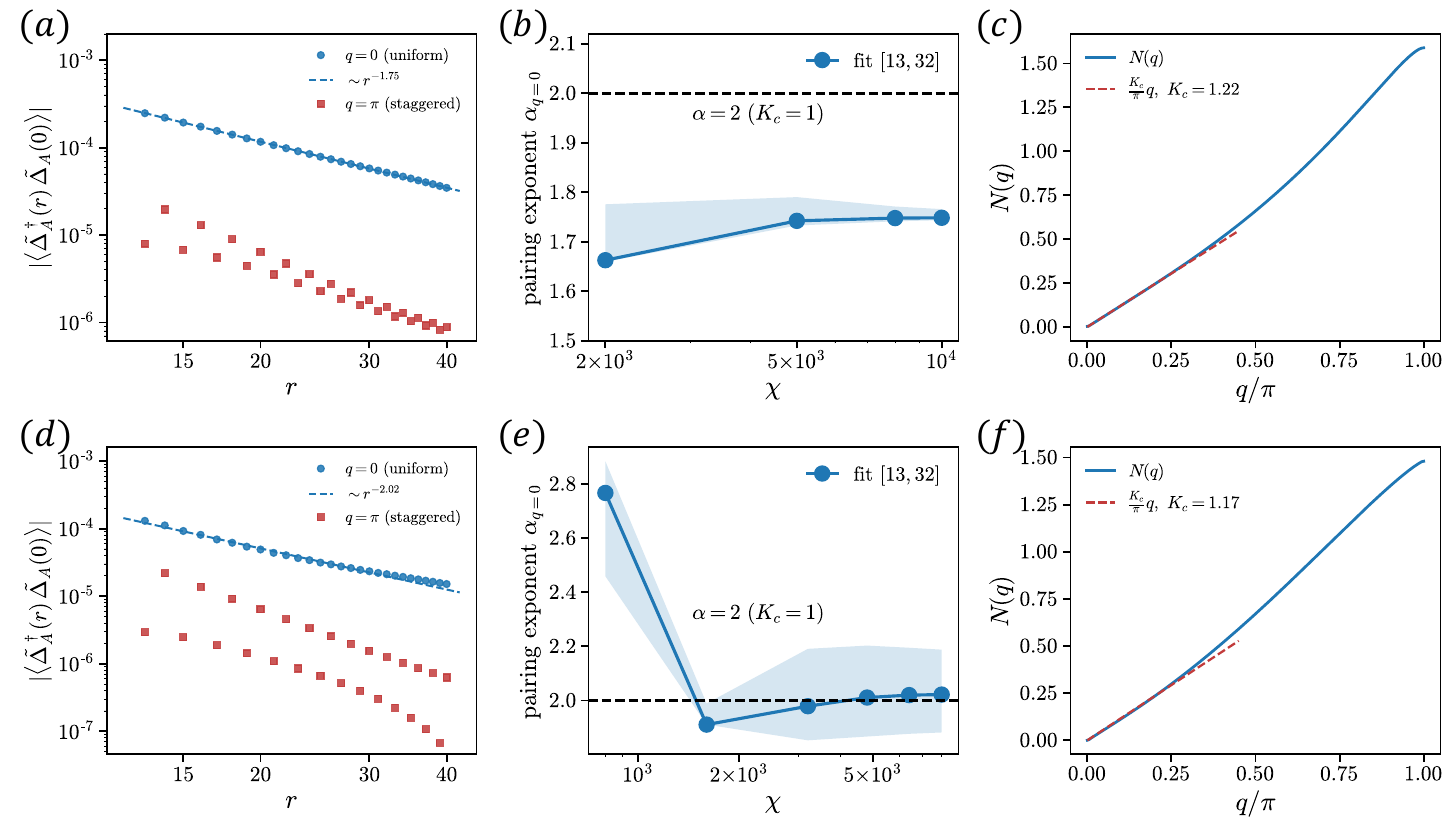}
\caption{
Bond-singlet pairing of the $\tilde{f}_A$,
$\tilde{\Delta}_A(i)=\tfrac{1}{\sqrt{2}}\left(\tilde f_{A;i;\uparrow}\tilde f_{A;i+1;\downarrow}
-\tilde f_{A;i;\downarrow}\tilde f_{A;i+1;\uparrow}\right)$.
(a)~$|\langle\tilde{\Delta}_A^\dagger(r)\tilde{\Delta}_A(0)\rangle|$ at $k_*=\infty,U_A=1,J_A=\frac{4}{3},U=\frac{1}{30},J=0$
and $\chi=10000$, decomposed into uniform and staggered components,
 the dashed line is a power-law fit over
$r\in[13,32]$. (b) Fitted exponent $\alpha_{q=0}$ versus bond dimension $\chi$, the shaded band
spans the fit windows $[13,25]$, $[13,32]$, and $[13,40]$. (c) Charge sturcture factor $N(q)$, which is $\frac{K_c}{\pi}|q|$ near $q=0$. (d)-(f):
Similar to (a)-(c), but use parameters $k_*=0.5,U_A=1,J_A=0.5,U=J=0$.}
\label{fig:pairing_exponent}
\end{figure}
In particular, in our calculations we can only estimate the value of $K_c$. The pairing operator in C1S2 phase is
\begin{equation}
    \Delta_{\mathrm{singlet},RL}\sim e^{\mathrm{i}\sqrt{2}\theta_c}\cos\left(\sqrt{2}\phi_s\right), \quad \Delta_{\mathrm{singlet},RR}\sim e^{\mathrm{i}\sqrt{2}(\theta_c-\phi_c)},
\end{equation}
where $RL,RR$ refers to the spin singlet formed by $R,L$ and $RR$ respectively. Similarly we can analyze $\Delta_{\mathrm{singlet},LL}$. Since we have
\begin{equation}
    \Delta [\Delta_{\mathrm{singlet},RL}]=\frac{1}{2}\left(1+\frac{1}{K_c}\right),\quad \Delta [\Delta_{\mathrm{singlet},RR}]=\frac{1}{2}\left(K_c+\frac{1}{K_c}\right),
\end{equation}
when $K_c>1$, the leading contribution to the pairing correlation function arises from $\Delta^\dagger_{\mathrm{singlet},RL}(x)\Delta_{\mathrm{singlet},RL}(0)$. Its long-distance behavior is $r^{-\alpha}$ with $\alpha=1+\frac{1}{K_c}<2$. Conversely, an exponent $\alpha>2$ would imply $K_c<1$. Our numerics clearly yields $\alpha<2$ in the $k_*=\infty$ case. For $k_*=0.5$ we have $\alpha\approx2$. Moreover, we calculate the charge sturcture factor $N(q)$, which is defined as the fourier transform of the density correlation function $\langle n(x)n(0)\rangle_c$. It scales as $N(q)\approx\frac{K_c}{\pi}|q|$ near $q=0$. We find $K_c>1$ for both $k_*=\infty$ and $k_*=0.5$. The results are shown in Fig.~\ref{fig:pairing_exponent}.


\end{document}